\documentclass[final,3p,number,sort&compress]{elsarticle}
\pdfoutput=1

\usepackage[english]{babel}
\usepackage[utf8]{inputenc}
\usepackage[T1]{fontenc}
\usepackage{charter}
\usepackage{setspace}

\graphicspath{{./figures/}}
\usepackage{graphicx,calc}
\usepackage[export]{adjustbox}
\usepackage{color,xcolor}
\definecolor{grau}{HTML}{6F6F6F}

\usepackage{tikz}
\tikzstyle{line}=[draw, thick, -stealth]

\usepackage[font=footnotesize]{caption}
\usepackage[font=footnotesize]{subcaption}
\usepackage{float}
\usepackage{listliketab}

\usepackage{tabularx}
\usepackage{array}
\usepackage{multirow}
\usepackage{booktabs}

\usepackage{amsmath}
\usepackage{amssymb}
\usepackage{amsfonts}
\usepackage{amsxtra}
\usepackage{mathtools}
\usepackage{empheq}
\usepackage{stmaryrd}
\usepackage{icomma}
\usepackage{relsize}

\usepackage[colorlinks=false,hidelinks]{hyperref}

\renewcommand\fbox{\fcolorbox{white!0}{white!0}} 
\newcolumntype{Y}{>{\raggedright\arraybackslash}X}
\newcolumntype{L}[1]{>{\raggedright\let\newline\\\arraybackslash\hspace{0pt}}p{#1}}
\newcolumntype{P}[1]{>{\centering\arraybackslash}m{#1}}
\newcommand{\eqn}[2]{\begin{equation} \label{#1} {#2} \end{equation}}
\newcommand{\grad}{\boldsymbol{\nabla}}
\newcommand{\bs}[1]{\boldsymbol{#1}}
\newcommand{\Vdir}{\text{\textsf{V}}\,}
\newcommand{\Tdir}{\text{\textsf{T}}\,}
\newcommand{\Hdir}{\text{\textsf{H}}\,}
\newcommand{\sdash}{\mbox{-}}

\journal{Engineering Fracture Mechanics (published version: \href{https://doi.org/10.1016/j.engfracmech.2024.110319}{doi.org/10.1016/j.engfracmech.2024.110319})} 

\title{Calibration and Validation of a Phase-Field Model of Brittle Fracture within the Damage Mechanics Challenge}

\author{Jonas~Heinzmann} \ead{jheinzmann@ethz.ch}
\author{Pietro~Carrara} \ead{pcarrara@ethz.ch}
\author{Chenyi~Luo} \ead{cheluo@ethz.ch}
\author{Manav Manav} \ead{mmanav@ethz.ch}
\author{Akanksha~Mishra} \ead{amishra@ethz.ch}
\author{Sindhu~Nagaraja} \ead{snagaraja@ethz.ch}
\author{Hamza~Oudich} \ead{houdich@ethz.ch}
\author{Francesco~Vicentini} \ead{fvicentini@ethz.ch}
\author{Laura~De~Lorenzis\corref{corauthor}} \ead{ldelorenzis@ethz.ch}
\cortext[corauthor]{Corresponding author, Email address: \url{ldelorenzis@ethz.ch}}

\address{Department of Mechanical and Process Engineering, ETH Z\"urich, Tannenstrasse 3, 8092 Zurich, Switzerland}

\begin{document}

\begin{frontmatter}

    \begin{abstract}
        In the context of the Damage Mechanics Challenge, we adopt a phase-field model of brittle fracture to blindly predict the behavior up to failure of a notched three-point-bending specimen loaded under mixed-mode conditions. The beam is additively manufactured using a geo-architected gypsum based on the combination of bassanite and a water-based binder. The calibration of the material parameters involved in the model is based on a set of available independent experimental tests and on a two-stage procedure. In the first stage an estimate of most of the elastic parameters is obtained, whereas the remaining parameters are optimized in the second stage to minimize the discrepancy between the numerical predictions and a set of experimental results on notched three-point-bending beams. The good agreement between numerical predictions and experimental results in terms of load-displacement curves and crack paths demonstrates the predictive ability of the model and the reliability of the calibration procedure.
    \end{abstract}

    \begin{highlights}
        \item The phase-field approach is used to model brittle fracture of additively manufactured orthotropic rock.
        \item We propose a calibration procedure based on a set of independent experimental tests.
        \item The behavior of an additional test is blindly predicted using the calibrated parameters.
        \item Experimental and numerical load--displacement curves and crack paths are compared.
        \item The fracture behavior is studied for different degrees of mode mixity.
    \end{highlights}


\end{frontmatter}

\section{Introduction}\label{sec:introduction}
The damage mechanics challenge (DMC), issued during a workshop held at Purdue University in 2019 \cite{dmc_agu2019}, aims at comparing the performances of different computational approaches and calibration strategies in predicting the behavior up to failure of a notched beam subjected to a mixed-mode three-point-bending (TPB) test \cite{dmc_challenge}.
The beam is made of a \textit{geo-architected} layered gypsum created using an additive manufacturing process \cite{dmc_material,dmc_material_supplementary} based on the crystallization of bassanite powder by means of a proprietary water-based binder.
While the experimental data for the \textsf{DMC} test geometry are not provided to the participants before the submission of their predictions, a set of independent characterization test results is made available \cite{dmc_calibration2,dmc_calibration1}.

This paper presents the modeling approach, the calibration strategy and the numerical results obtained by the GRIPHFiTH team from the Computational Mechanics Group of ETH Z\"urich.
We adopt the phase-field approach to brittle fracture, originally proposed by Bourdin et al. \cite{bourdin_numerical_2000} as the regularization of the variational reformulation of Griffith's brittle fracture in \cite{francfort_revisiting_1998}, and later re-interpreted as a special class of gradient damage models \cite{pham_gradient_2011}. 
In models of this type, beside the displacement field, an additional field variable, denoted as phase-field or damage variable $d$, smoothly varies between 0 and 1 to denote the transition of the material state between pristine and fully broken conditions, respectively.
The numerical implementation of the approach does not require remeshing upon crack propagation nor a priori knowledge of the crack path; crack nucleation, branching, merging and any other topological changes are automatically described with no need for ad-hoc criteria.
These features, along with a strong mathematical background, make the phase-field approach very attractive to model various fracture phenomena including brittle \cite{bourdin_numerical_2000,pham_onset_2013,tanne_crack_2018}, ductile \cite{ambati_phase-field_2015,miehe_phase_2016,alessi_gradient_2015}, anisotropic \cite{clayton_phase_2014,teichtmeister_phase_2017,vandijk_strain_2020,nagaraja_deterministic_2023,li_crack_2019} and fatigue fracture \cite{Carrara2020,Seiler2020,mesgarnejad_phase-field_2019,boldrini_non-isothermal_2016}, as well as fracture in heterogeneous materials \cite{hansen-dorr_phase-field_2021,vicentini_phase-field_2023,hossain_effective_2014}.

In this paper, we base our predictions on the assumption of a macroscopically homogeneous orthotropic material undergoing brittle fracture and we calibrate the material parameters following a procedure subdivided in two stages. In the first stage, we obtain an estimate for most of the elastic parameters using the results of plane wave propagation velocity tests and unconfined compression tests, part of the available DMC data. This allows us to limit the number of free parameters to be optimized during the second stage to two, namely one elastic parameter and the fracture toughness. The optimization procedure of the second stage aims at minimizing the mismatch between the experimental load-deflection curves from a set of notched TPB tests, also part of the available DMC data, and their numerical prediction, as measured by a properly defined cost function.
The adopted model and the calibrated parameters are first verified by comparing the numerical and the experimental results for the same notched TPB specimens used for calibration, in terms of both load-deflection curves and crack paths. Then, the behavior of the specimen with the \textsf{DMC} test geometry is blindly predicted; finally, the corresponding numerical results are compared to the experimental results made available by the organizers after submission of the team predictions.  

This paper is structured as follows:
Section~\ref{sec:overview_of_experiments_and_modeling} briefly overviews the DMC tasks and the experimental data available for model calibration.
Section~\ref{sec:modeling_and_numerical_aspects} illustrates the adopted model and the related numerical aspects, while Section \ref{sec:parameter_calibration} is devoted to the calibration of the material parameters.
Section~\ref{sec:results_and_discussion} presents the comparison between numerical and experimental data for the notched TPB tests used for calibration and for the \textsf{DMC} test geometry. Conclusions are drawn in Section~\ref{sec:conclusions}.

\section{Damage mechanics challenge}\label{sec:overview_of_experiments_and_modeling}
In this section, we summarize the main tasks assigned to the DMC participants and the characterization tests used to calibrate the adopted model.
For a more detailed description, the interested reader should refer to \cite{dmc_qa}.

\subsection{Damage mechanics challenge task}\label{sec:DMC_task}
\begin{figure}
    \centering
    \subfloat[]{
        {\raisebox{4.25cm}{\textsf{DMC}}\hspace*{-2.0cm}}\includegraphics{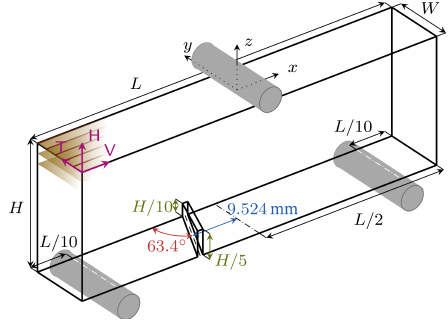}
        \label{fig:geometry_HCH}
    }
    \subfloat[]{
        \begin{tabular}[b]{cc}
            {\raisebox{2.25cm}{\textsf{HC}}\hspace*{-0.75cm}}\includegraphics{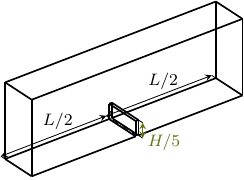} &{\raisebox{2.25cm}{\textsf{HB}}\hspace*{-0.75cm}}\includegraphics{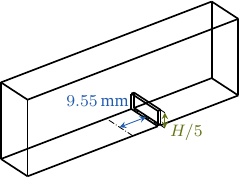}\\
            {\raisebox{2.25cm}{\textsf{HA}}\hspace*{-0.75cm}}\includegraphics{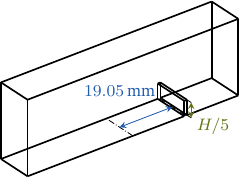} &{\raisebox{2.25cm}{\textsf{H45}}\hspace*{-0.75cm}}\includegraphics{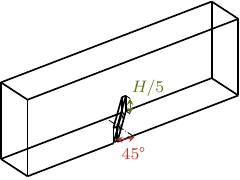}
        \end{tabular}\hspace*{-1cm}
        \label{fig:geometry_HABC45}}
    \caption{Geometry and loading conditions of the notched beams for (a) the \textsf{DMC} test and (b) the characterization tests. All the beams have the same width $W =$ 12.7~mm, height $H =$ 25.4~mm and length $L =$ 76.2~mm; the midspan is indicated with a dashed-dotted line.}
    \label{fig:geometries}
\end{figure}

The main assigned task involves the prediction of the behavior up to complete failure of a notched beam made of an additively manufactured geomaterial subjected to a monotonic TPB test, whose geometry and loading conditions are summarized in Fig.~\ref{fig:geometry_HCH}.
The beam is supported and loaded by means of three aluminum rods covering its complete width.
In particular, the two lower rods act as supports, while the third one transmits the load to the midspan section of the beam.
The test is performed in displacement control by imposing the midspan deflection at a rate of 0.03~mm/min, hence sufficiently slow to consider the test as quasi-static.
The notch is directly included in the specimen geometry during the manufacturing process and has a width of 1~mm, is off-centered by 9.52~mm with respect to the midspan position, and its longitudinal axis forms an angle of about 63.4$^\circ$ with the longitudinal axis of the beam.
Also, its height is linearly tapered and decreases from 5.08~mm to 2.54~mm across the width (Fig. \ref{fig:geometry_HCH}). This setup aims at inducing a complex mixed-mode fracture condition including all three modes, namely opening or mode I, sliding or mode II and tearing or mode III.

The complexity of the assigned task stems not only from the mode mixity but also from the material features. The beam is obtained through an additive manufacturing process, in which layers of bassanite (i.e., calcium sulfate hemi-hydrate) powder with a thickness of about 0.1~mm are bonded using a proprietary water-based binder that, reacting with the bassanite, produces gypsum as by-product \cite{dmc_slides,dmc_material,dmc_material_supplementary}.
The binder is sprayed on the bassanite layer in bands with a width of about 0.5~mm through a nozzle, similarly as in an ink-jet printer, creating the so-called \textit{mineral bands} \cite{dmc_material}.
The resulting material is clearly heterogeneous and anisotropic for different reasons: firstly, the binding of the bassanite powder leaves pores at the mineral bands scale.
Secondly, the deposition of the binder induces a preferential growth of the gypsum crystals along the spraying direction and an irregular distribution of bonded material at the interfaces between adjacent mineral bands and between successive depositions of bassanite layers (also termed in short \textit{deposition layers}) \cite{dmc_material}.
The presence of interfaces and of a preferential direction for the growth of the gypsum crystals induces a macroscopic material behavior typical of \textit{orthorombic} systems, namely an \textit{orthotropic} behavior \cite{dmc_material}.
In the adopted coordinate system, the bassanite layers are deposited in the $x\mbox{-}y$ plane (brown planes in Fig.~\ref{fig:geometry_HCH}), while the binder is sprayed along the $x$-direction (brown lines in Fig.~\ref{fig:geometry_HCH}), which is thus the direction of the mineral bands.
To facilitate the identification of the directions related to the manufacturing process we define here the \textit{material directions} as
\begin{itemize}[-]
    \item \textsf{\textbf{V}}\textbf{-direction}: binder spraying or mineral bands direction, corresponding to the $x$-direction;
    \item \textsf{\textbf{T}}\textbf{-direction}: direction within the deposition layer plane and orthogonal to the mineral bands, corresponding to the $y$-direction; and
    \item \textsf{\textbf{H}}\textbf{-direction}: direction orthogonal to the deposition layer plane, corresponding to the $z$-direction,
\end{itemize}
as indicated in Fig.~\ref{fig:geometry_HCH}.

\subsection{Experimental characterization tests}
To characterize the material behavior, an extensive set of experimental data is provided in \cite{dmc_calibration1,dmc_calibration2,dmc_slides}. For the sake of clarity we summarize here only the data used in the following, which include
\begin{itemize}[-]
    \item load-displacement, digital image correlation and crack surface roughness data of TPB tests on beams differing from the \textsf{DMC} test beam only by the notch geometry\footnote{The outer dimensions of the beams and the positions of the support and loading rods are the same as in Fig.~\ref{fig:geometry_HCH}.}.
    These beams comprise (Fig. \ref{fig:geometry_HABC45})
          \begin{enumerate}[i.]
              \item four specimens labeled as \textsf{HC}, designed to induce a crack under mode-I conditions;
              \item four specimens labeled as \textsf{HB}, designed to induce a crack under mixed-mode (I+II) conditions with mode-I predominance;
              \item four specimens labeled as \textsf{HA}, designed to induce a crack under mixed-mode (I+II) conditions with mode-II predominance;
              \item four specimens labeled as \textsf{H45}, designed to induce a crack under mixed-mode (I+III) conditions;
          \end{enumerate}
    \item five load-displacement curves for each material orientation obtained from unconfined compression tests;
    \item density measurements on three cubic samples;
    \item longitudinal and shear wave velocities from ultrasonic measurements for material directions \textsf{V}, \textsf{T} and \textsf{H}.
\end{itemize}

\section{Modeling and computational aspects}\label{sec:modeling_and_numerical_aspects}
This section briefly summarizes the model and the computational strategy used to carry out the DMC task assignment.

\subsection{Phase-field model of brittle fracture}\label{sct:model}
In the following, we assume linearized kinematics, rate independence, quasi-static and isothermal conditions and negligible body forces.
Moreover, since the dimensions of the heterogeneities generated by the manufacturing process are much smaller than the specimen dimensions, we assume a homogeneous material featuring orthotropic elasticity and fracture.

Let us consider a body $\Omega$ with boundary $\partial \Omega$. Introducing the spatial coordinate $\bs{x} \in \Omega$, we denote the displacement field as $\bs{u}(\bs{x})$ and the stress and strain fields respectively as  $\bs{\sigma}(\bs{x})$ and $\bs{\varepsilon}(\bs{x})=\grad_{\text{sym}}(\bs{u}(\bs{x}))$, where $\grad_{\text{sym}}(\bullet)$ is the symmetric gradient of $(\bullet)$.
Boundary tractions $\bs{\bar t}(\bs{x})$ are prescribed on the Neumann part of the boundary $\partial \Omega_{\text{N}}$ and imposed displacements $\bs{\bar u}(\bs{x})$ are applied to the Dirichlet part of the boundary  $\partial \Omega_{\text{D}}$, with  $\partial \Omega = \partial \Omega_{\text{N}} \cup \partial \Omega_{\text{D}}$, $\partial \Omega_{\text{N}} \cap \partial \Omega_{\text{D}}=\varnothing$.
We also define $\bs{n}$ as the outward unit normal to $\partial \Omega$.
\begin{figure}
    \centering
    \subfloat[]{
        \hspace{-0.4cm}
        \includegraphics{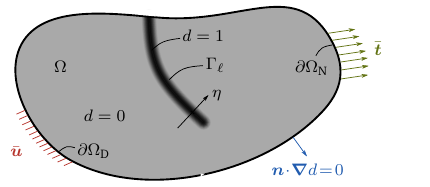}
        \label{fig:fractured_domain_regularized}
    }
    \subfloat[]{
        \includegraphics{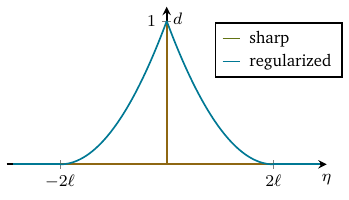}
        \label{fig:at1_profile}
    }
    \caption{(a) Schematic representation of the domain $\Omega$ with prescribed traction $\bar{\bs{t}}$ on the Neumann boundary $\partial \Omega_{\text{N}}$ and prescribed displacement $\bar{\bs{u}}$ on the Dirichlet boundary $\partial \Omega_{\text{D}}$.
        The optimal profile of the phase-field variable along a coordinate perpendicular to the axis of the regularized crack $\Gamma_\ell$ is shown in (b) in comparison to the sharp crack representation which is recovered for a length-scale parameter $\ell \rightarrow 0$.}
    \label{fig:crack_regularization}
\end{figure}

To model the material behavior we adopt a \textit{phase-field model of brittle fracture}, whereby damage and fracture are described through a \textit{phase-field} or \textit{damage variable} $d$; this variable features a steep but smooth transition from $d=0$, corresponding to an intact material, to $d=1$, denoting a fully damaged material, over a support whose size is governed by a length-scale parameter $\ell$ \cite{Ambati2014} (Fig.~\ref{fig:crack_regularization}).
The phase-field approach thus smears a sharp crack over a small but not vanishing portion of the domain, $\Gamma_{\ell}$ (Fig.~\ref{fig:fractured_domain_regularized}).

The governing equations of the model are derived in a variational fashion from the local minimization of the total energy functional, which reads
\begin{equation}
    \mathcal{E} (\bs{u}, d) = \int_\Omega \underbrace{g(d) \psi_0 \left(\bs{\varepsilon}\right) \vphantom{\frac{\mathcal{G}_{\text{c}}}{4 c_w} \left( \frac{w(d)}{\ell} + \ell |\grad d|^2 \right)}}_{\displaystyle\psi\left(\bs{\varepsilon},\,d\right)}\mathrm{d}V + \int_\Omega \underbrace{\frac{\mathcal{G}_{\text{c}}}{4 c_w} \left( \frac{w(d)}{\ell} + \ell |\grad d|^2 \right)}_{\displaystyle\mathcal{D}\left(d\right)}\mathrm{d}V - \int_{\partial \Omega_N}\vphantom{\frac{\mathcal{G}_{\text{c}}}{4 c_w} \left( \frac{w(d)}{\ell} + \ell |\grad d|^2 \right)}\bs{\bar t} \cdot \bs{u} \, \mathrm{d}A \text{ .}
    \label{eq:energy_functional}
\end{equation}
The strain energy density $\psi\left(\bs{\varepsilon},\,d\right)$ depends on the monotonically decreasing scalar function $g(d)= (1-d)^2 + g_0$, termed \textit{degradation function}, where the parameter $g_0 = o(\ell)$ (here chosen as $10^{-5}$) provides a small residual stiffness preventing numerical instabilities \cite{bourdin_numerical_2000}.
A comparison of different degradation functions is given in \cite{Kuhn_2015_degradation}.
This function governs how the material transitions between intact and broken states and is directly applied to the undegraded elastic strain energy density $\psi_0\left(\bs{\varepsilon}\right)$, which reads
\begin{equation}
    \psi_0(\bs{\varepsilon}) = \frac{1}{2} \boldsymbol{\varepsilon} : \mathbb{C} : \boldsymbol{\varepsilon} \text{ .}
\end{equation}
As a consequence of the assumed orthotropic elastic behavior, the symmetric fourth-order elasticity tensor $\mathbb{C}$ contains nine independent components  and, in Voigt notation, reads
\begin{equation}
    \left[ \mathbb{C} \right] =
    \begin{bmatrix}
        C_{xxxx} & C_{xxyy} & C_{xxzz} & 0        & 0        & 0        \\
        C_{xxyy} & C_{yyyy} & C_{yyzz} & 0        & 0        & 0        \\
        C_{xxzz} & C_{yyzz} & C_{zzzz} & 0        & 0        & 0        \\
        0        & 0        & 0        & C_{yzzy} & 0        & 0        \\
        0        & 0        & 0        & 0        & C_{xzzx} & 0        \\
        0        & 0        & 0        & 0        & 0        & C_{xyyx} \\
    \end{bmatrix} \text{ .}
\end{equation}
The dissipated energy density $\mathcal{D}(d)$ in \eqref{eq:energy_functional} depends not only on the phase-field variable but also on its gradient $\grad d$. The normalization constant $c_w$ ensures that the energy dissipated per unit length of a regularized crack corresponds to the fracture toughness $\mathcal{G}_{\text{c}}$.
Its value depends on the choice of the monotonically increasing \textit{dissipation function} $w(d)$, which modulates the energy dissipation with the increase of homogeneous damage.
Although alternatives are available \cite{Lorentz_convergence_2019, alessi_gradient_2014}, the two most common dissipation functions used for brittle materials are $w(d) = d$ (\textsf{AT1} model), yielding $c_w=2/3$,  and $w(d)=d^2$ (\textsf{AT2} model), leading to $c_w=1/2$.
The main difference between the two is that the former leads to an initial linear elastic stage of behavior, whereas with the latter damage starts evolving already at vanishing stress \cite{pham_gradient_2011}.

The functional presented in \eqref{eq:energy_functional} yields a symmetric evolution of the phase-field variable under tensile and compressive stress states.
This implies that a crack can nucleate and propagate also under mainly compressive states, and that the faces of a fully developed crack can exhibit interpenetration upon closure.
To overcome these issues, different approaches are available in the literature; they are mostly based on an additive split of the undegraded elastic strain energy into an \textit{active} and an \textit{inactive} part \cite{miehe_phase_2010,amor_regularized_2009,freddi_regularized_2010,de_lorenzis_nucleation_2022,Vicentini2023}.
While this split is essential to obtain meaningful results in problems with significant compression \cite{Vicentini2023}, in cases with predominant tension it does not significantly modify results with respect to the symmetric model, and at the same time it leads to an increased computational cost.
Therefore, since here we focus on notched TPB tests for which predominant tension is anticipated, we keep the symmetric model based on the total energy \eqref{eq:energy_functional}. A further simplifying assumption inherent to \eqref{eq:energy_functional} is that the orthotropic behavior determined by the manufacturing process is limited to the elastic stiffness, whereas the damage behavior stays isotropic. Although phase-field models addressing a directional dependence of $\mathcal{G}_{\text{c}}$ have been presented in the literature \cite{teichtmeister_phase_2017,nagaraja_deterministic_2023}, our reasons for this assumption are that the degree of material anisotropy is found to be moderate (see Section \ref{sct:el_est}), and that the available data are insufficient to properly characterize an anisotropic fracture model.

The governing equations are derived by applying the principles of irreversibility, energy balance and first-order stability \cite{pham_gradient_2011}, leading to the equations summarized in Tab.~\ref{tab:probl}.
For the phase-field variable, beside the Neumann boundary conditions that naturally emerge from the variational formulation, we introduce in \eqref{eq:pf_BC} an additional Dirichlet condition, denoted as internal condition, in order to prevent the evolution of spurious damage in the subset $\Omega_{\text{P}}$ of the domain. 
As further detailed later, $\Omega_{\text{P}}$ consists of the portions of the domain close to the support and loading rods. 

\begin{table}
    \caption{Governing equations and boundary conditions of the problem.}
    \label{tab:probl}
    \footnotesize
    \centering
    \begin{tabular}{P{3cm}P{12.2cm}}
        \toprule
        \multicolumn{2}{c}{\textbf{Linear momentum balance}}\\
        \toprule
        \textbf{\vspace{3mm}Equilibrium equation}             &\vspace{-5mm}\eqn{eq:equilibrium}{\grad \cdot \bs{\sigma}  = \bs{0}\, \ \  \forall \bs{x} \in \Omega\,}\\[-1em]
        \textbf{\vspace{3mm}Cauchy stress}                    &\vspace{-5mm}\eqn{eq:stress}{\bs{\sigma}(\bs{\varepsilon},d) = \frac{{\partial} \psi(\bs{\varepsilon},d)}{ {\partial} \bs{\varepsilon}}=g(d) \frac{{\partial} \psi_0(\bs{\varepsilon})}{ {\partial} \bs{\varepsilon}}}\\[-1em]
        \textbf{\vspace{3mm}Degradation function}             &\vspace{-5mm}\eqn{eq:degr_fct}{g(d)=(1-d)^2+g_0\,\quad \text{with} \quad g_0=10^{-5}}\\[-1em]
        \textbf{\vspace{3mm}Boundary conditions}              &\vspace{-5mm}\eqn{eq:equi_BC}{\begin{cases} \bs{\sigma} \cdot \bs{n} = \bar{\bs{t}} & \forall \bs{x} \in \partial \Omega_{\text{N}}\,\\ \bs{u} = \bar{\bs{u}}\,  & \forall \bs{x} \in \partial \Omega_{\text{D}}\,\end{cases}}\\[-1em]
        \textbf{\vspace{3mm}Parameters}                       &\multirow{2}{*}{\vspace{3mm}$C_{xxxx},\ C_{yyyy},\ C_{zzzz},\ C_{xxyy}, \ C_{yyzz},\ C_{zzxx},\ C_{xyyx}, \ C_{yzzy}, \ C_{zxxz} \to \mathbb{C}$  }\\[-1em]
        {\vspace{0mm}(Total: 9)}                              &\\[0em]
        \toprule
        \multicolumn{2}{c}{\textbf{Damage evolution}}\\
        \toprule
        \textbf{\vspace{3mm}KKT conditions}                   &\vspace{-5mm}\eqn{eq:phasefield}{\begin{cases}\dot{d} \geq 0\,\\
                                                                                                         \displaystyle g^\prime(d) \psi_0 + \frac{\mathcal{G}_{\text{c}}}{4c_w } \! \left(\frac{w^{\prime}(d)}{\ell} - 2 \ell \grad \! \cdot \! (\grad d) \right) \geq 0\, \\[1em]
                                                                                                         \displaystyle\left[ g^\prime(d) \psi_0 + \frac{\mathcal{G}_{\text{c}}}{4c_w } \! \left(\frac{w^{\prime}(d)}{\ell} - 2\ell \grad \! \cdot \! (\grad d) \right) \right] \dot{d} = 0\,
                                                                                                \end{cases}}\\[-.5em]
        \vspace{-2mm}\textbf{Boundary \& internal conditions} &\vspace{-5mm}\eqn{eq:pf_BC}{\begin{cases}\bs{n} \cdot \grad d= 0\,, & \forall \bs{x} \in \partial \Omega\setminus\left(\partial \Omega \cap \Omega_{\text{P}}\right)\,\\ d = 0\,  & \forall \bs{x} \in \Omega_{\text{P}}\,\end{cases}}                                   \\[-.5em]
        \textbf{\vspace{3mm}Dissipation function}             &\vspace{-5mm}\eqn{eq:diss_fct}{w(d)=d\quad \text{with}\quad c_w=\frac{2}{3} \ \ \textsf{(AT1)}\,,\quad \text{or} \quad w(d)=d^2\quad \text{with}\quad c_w=\frac{1}{2}\ \  \textsf{(AT2)}}\\[-.5em]
        \textbf{\vspace{3mm}Parameters}                       &\multirow{2}{*}{\vspace{3mm}$\ell,\ \mathcal{G}_{\text{c}}$}\\[-1em]
        {\vspace{0mm}(Total: 2)}                              &\\[0em]
        \toprule
    \end{tabular}
\end{table}

For the present modeling choices, in absence of stress singularities within the domain the length-scale parameter $\ell$ controls the material strength and can thus be regarded as a material parameter \cite{tanne_crack_2018}.
Conversely, in presence of stress singularities, results are expected to be fairly independent on $\ell$ as long as it is sufficiently small compared to the smallest characteristic length in the domain.
Thus, in such cases $\ell$ plays the role of a numerical parameter determining the thickness of the regularized representation of a crack \cite{tanne_crack_2018}.
However, if $\ell$ is comparable to the smallest characteristic length in the domain, the system response becomes dependent on its value, in which case $\ell$ should be treated as a further model parameter \cite{Mandal_length_2019}.

\subsection{Computational aspects}\label{sct:num_asp}
The solution of the non-linear system of coupled equations arising from Tab.~\ref{tab:probl} requires a spatial discretization method and an incremental iterative solution scheme.

For spatial discretization, we adopt the standard finite element (FE) method based on hexahedral 8-node FEs with four degrees of freedom (DOFs) per node, i.e. the three components of the displacement field $\bs{u}$ and the phase-field variable $d$.
The discretization of the domain must be sufficiently fine to properly resolve the length-scale parameter $\ell$.
As better illustrated in the following sections, FE computations are used for both the calibration of the model, based on the TPB tests on specimens \textsf{HC}, \textsf{HB}, \textsf{HA} and \textsf{H45}, and the prediction of the TPB test on the \textsf{DMC} specimen.
Since all these specimens feature a pre-existing notch, we set the length scale to $\ell$ = 0.625~mm, which is considered sufficiently small to treat it as a numerical parameter (Fig.~\ref{fig:geometries}) but still large enough to achieve a reasonable computational cost.

\begin{figure}[t]
    \centering
    \subfloat[]{
        \begin{tabular}[b]{c}
            {\footnotesize \textsf{HC} test}\\[-.5em]
            {\footnotesize  (notch detail, front, bottom view)}\\
            \includegraphics[width=3.5cm]{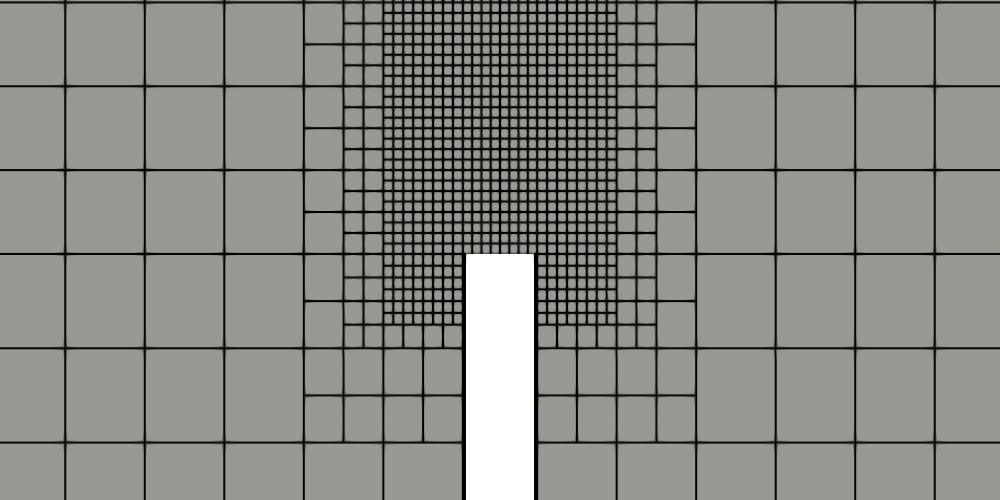}\\
            \includegraphics[width=5cm]{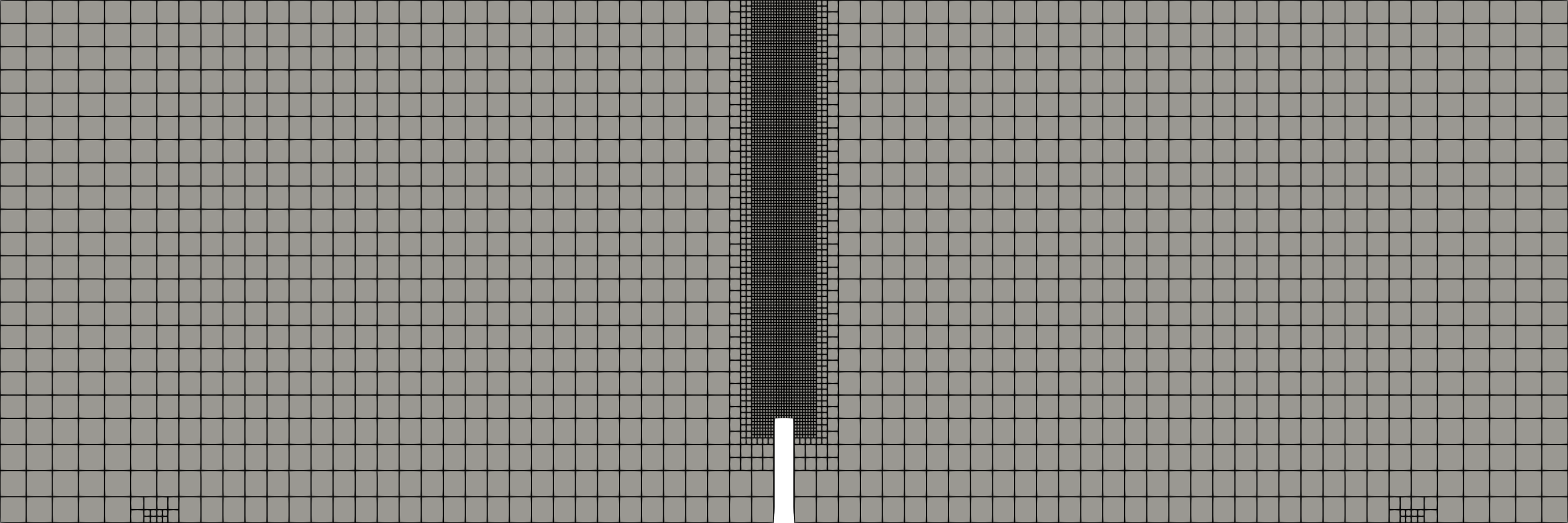}\\
            \includegraphics[width=5cm]{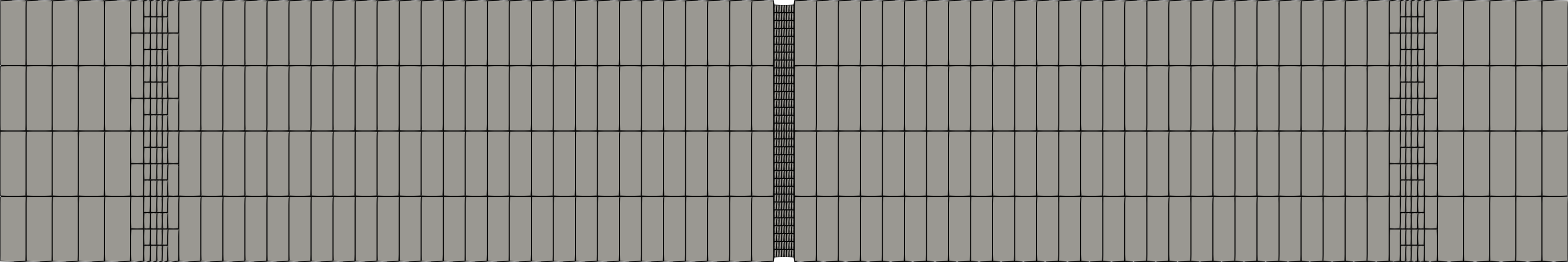}\\
        \end{tabular}\label{fig:meshHC}}
    \subfloat[]{
        \begin{tabular}[b]{c}
            {\footnotesize \textsf{HB} test}\\[-.5em]
            {\footnotesize  (notch detail, front, bottom view)}\\
            \includegraphics[width=3.5cm]{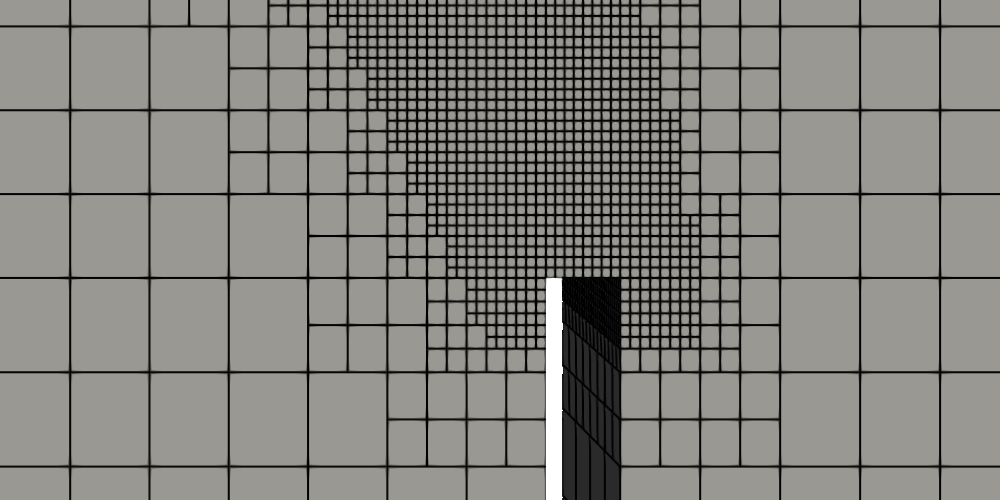}\\
            \includegraphics[width=5cm]{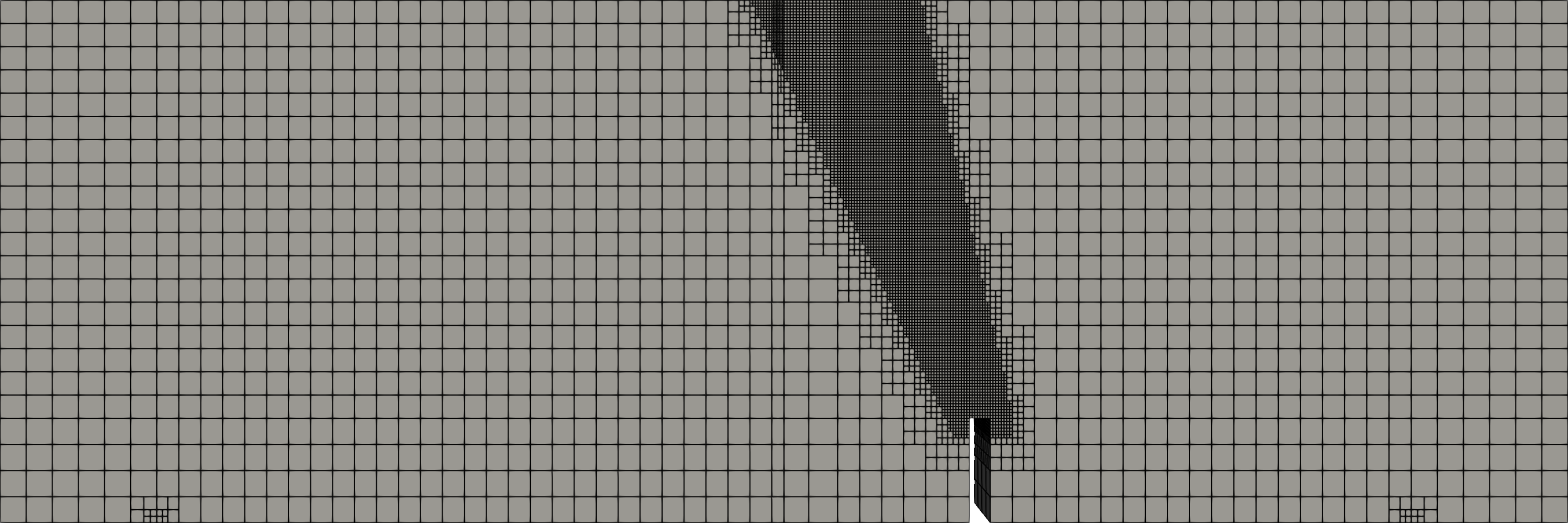}\\
            \includegraphics[width=5cm]{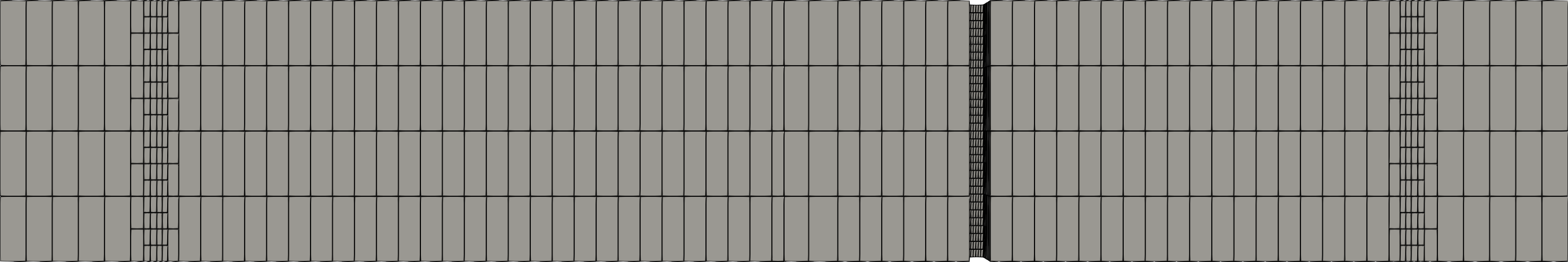}\\
        \end{tabular}\label{fig:meshHB}}
    \subfloat[]{
        \begin{tabular}[b]{c}
            {\footnotesize \textsf{HA} test}\\[-.5em]
            {\footnotesize  (notch detail, front, bottom view)}\\
            \includegraphics[width=3.5cm]{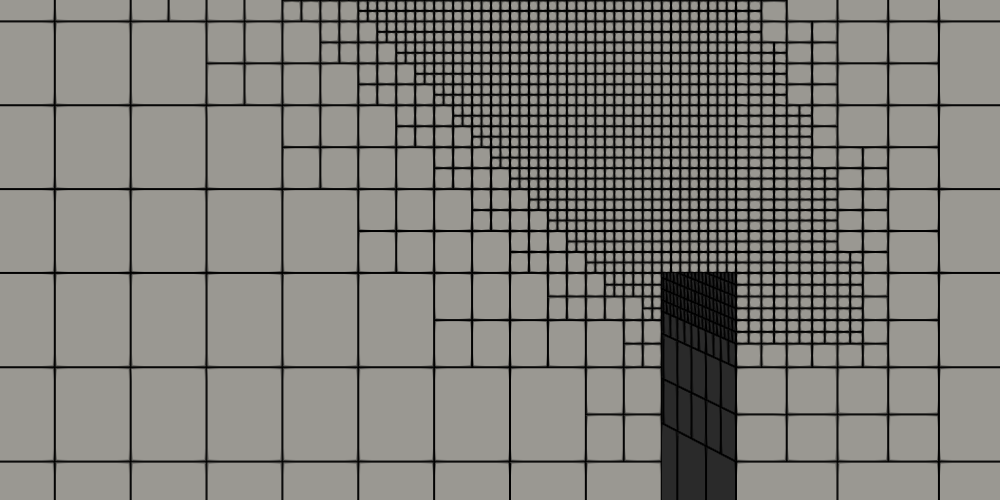}\\
            \includegraphics[width=5cm]{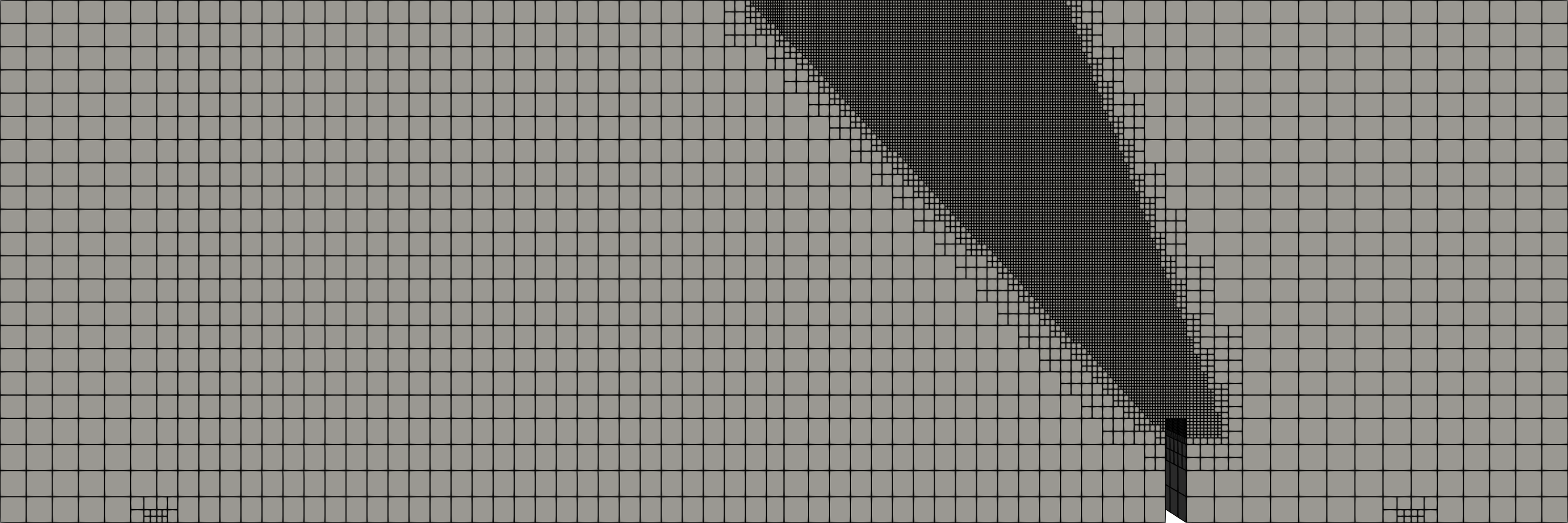}\\
            \includegraphics[width=5cm]{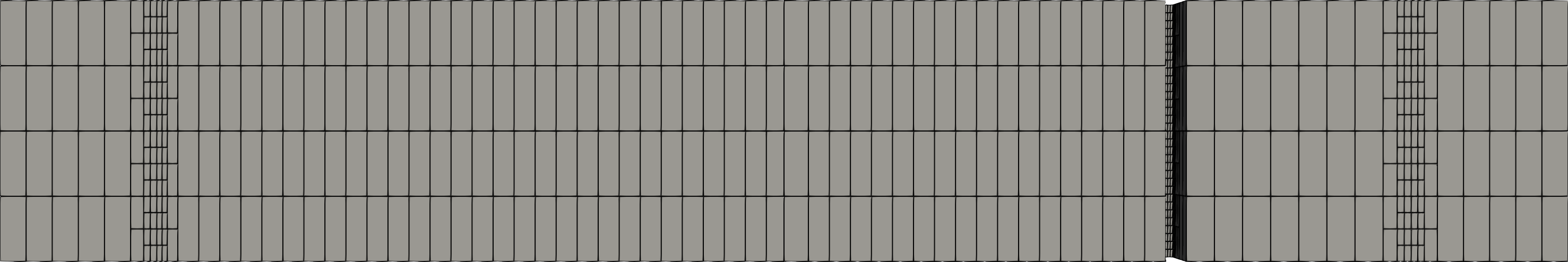}\\
        \end{tabular}\label{fig:meshHA}}
    \newline
    \subfloat[]{
        \begin{tabular}[b]{c}
            {\footnotesize \textsf{H45} test}\\[-.5em]
            {\footnotesize  (notch detail, front, bottom view)}\\
            \includegraphics[width=3.5cm]{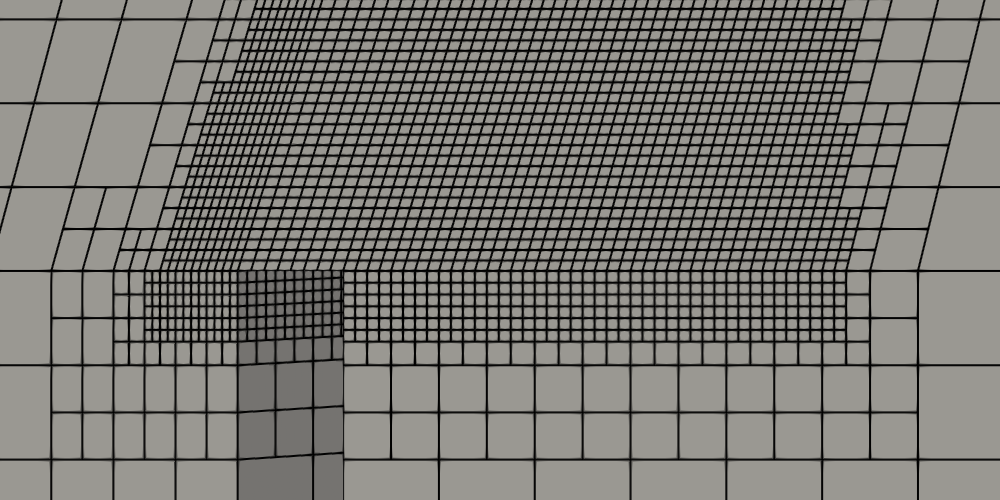}\\
            \includegraphics[width=5cm]{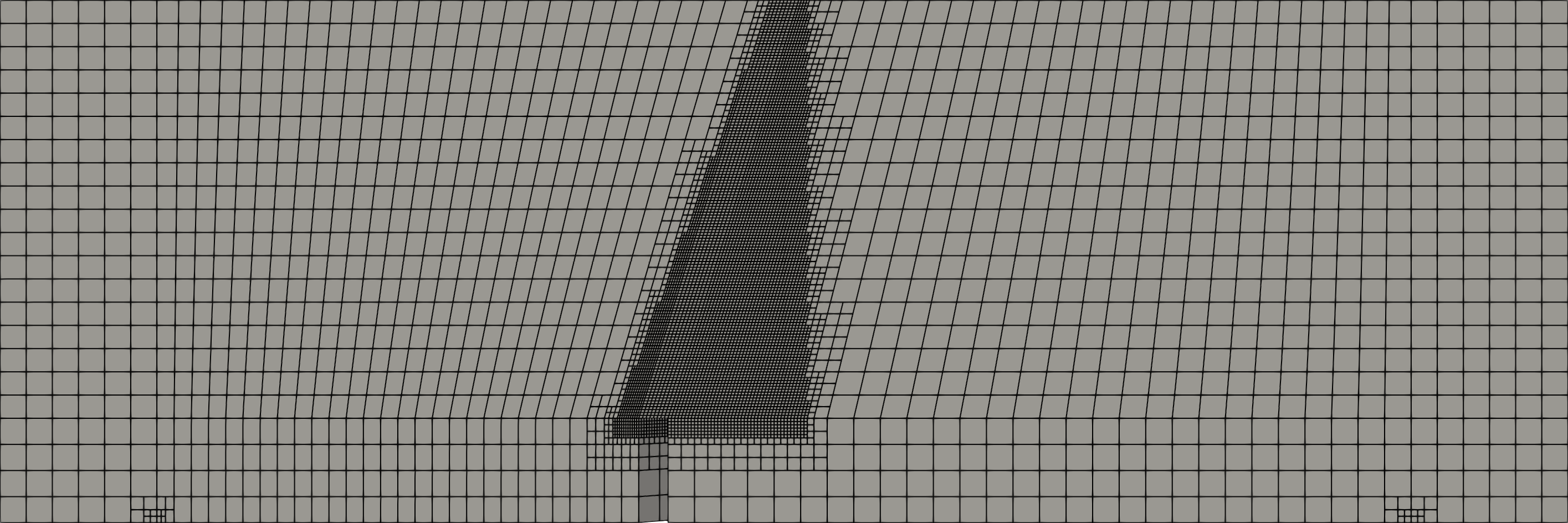}\\
            \includegraphics[width=5cm]{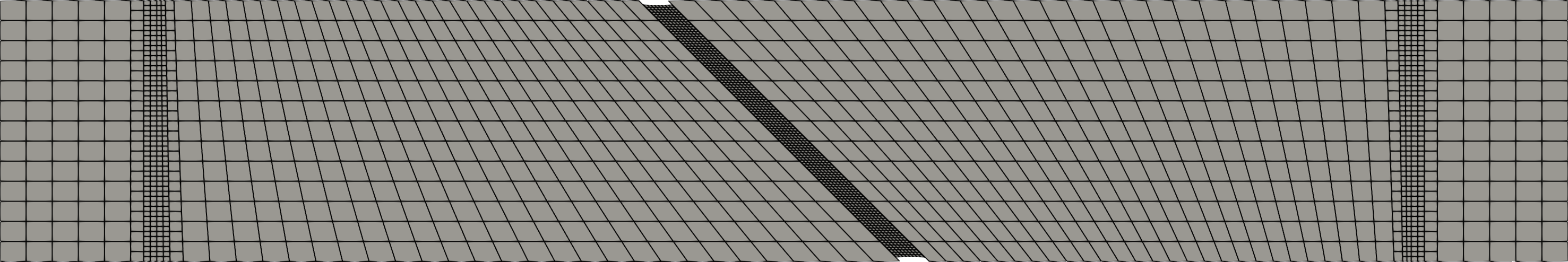}\\
        \end{tabular}\label{fig:meshH45}}
    \subfloat[]{
        \begin{tabular}[b]{cc}
            \multicolumn{2}{c}{\footnotesize \textsf{DMC} test}\\[-.5em]
            {\footnotesize  (notch detail \& front view)}        &{\footnotesize  (top, back, bottom view)}\\
            \includegraphics[width=3.5cm]{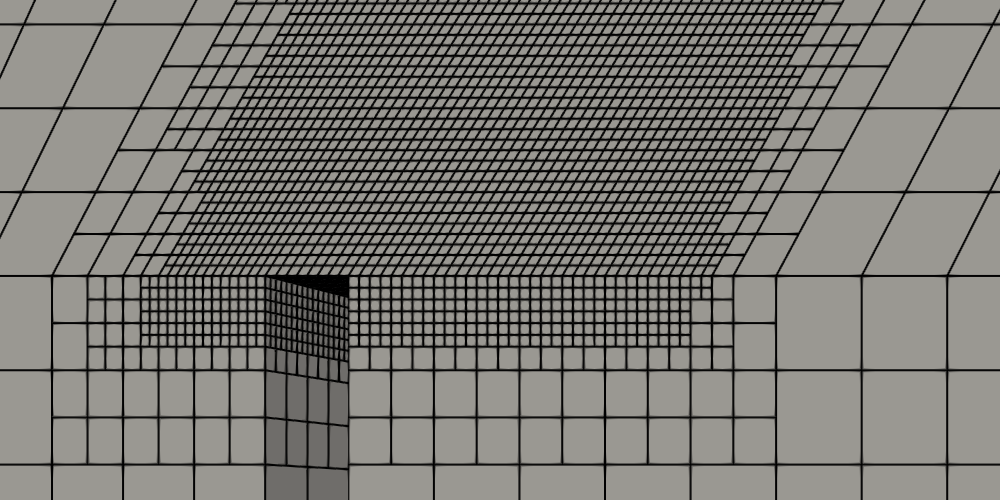}  &\hspace{0.12cm}\includegraphics[width=5cm]{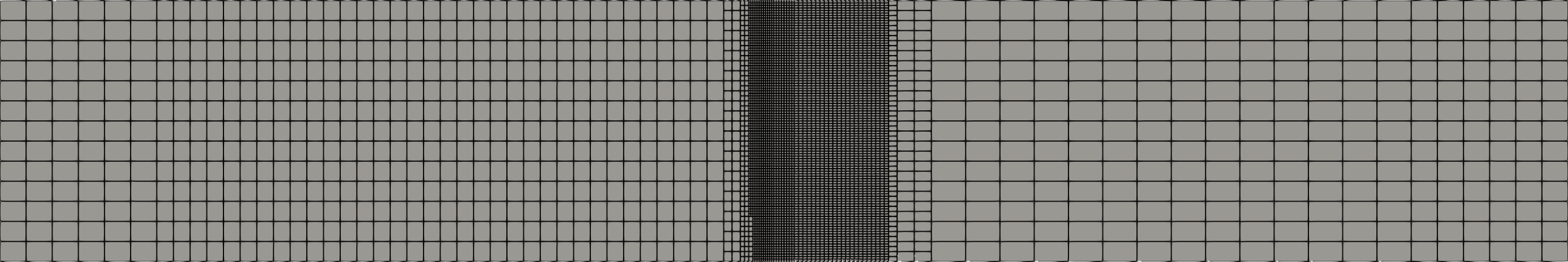}\\
            \includegraphics[width=5cm]{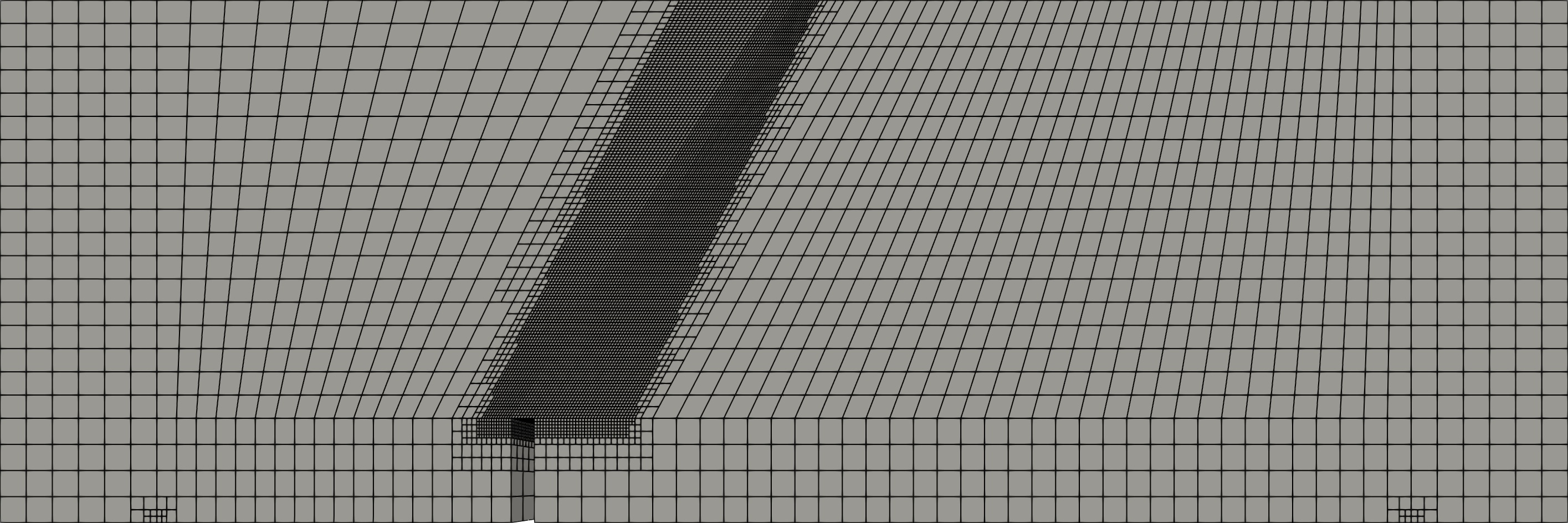}    &\hspace{0.12cm}\includegraphics[width=5cm]{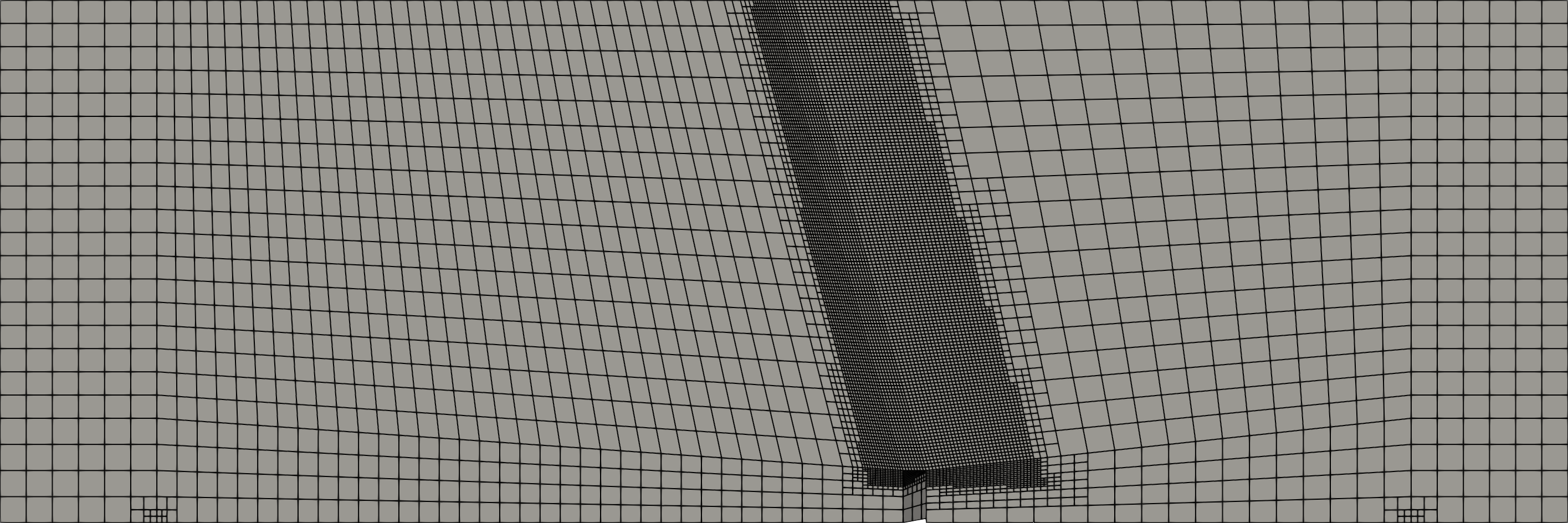}\\
                                                                 &\hspace{0.12cm}\includegraphics[width=5cm]{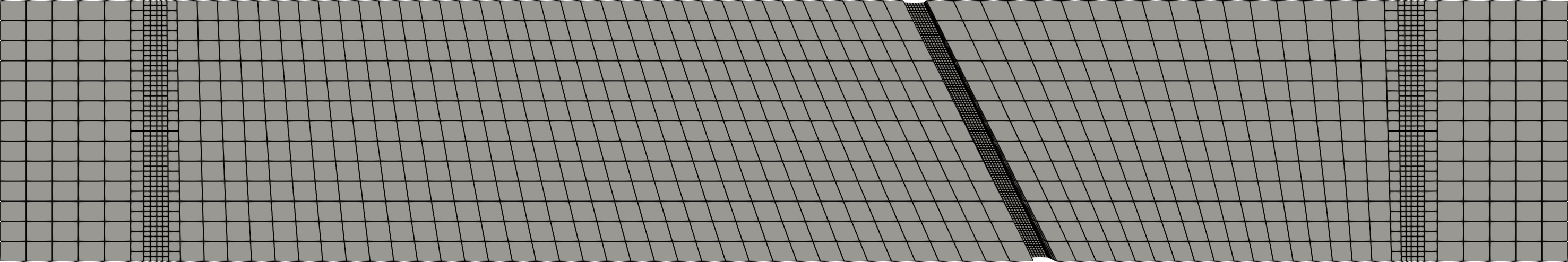}\\
        \end{tabular}\label{fig:meshDMC}}
    \caption{Meshes for the (a) \textsf{HC}, (b) \textsf{HB}, (c) \textsf{HA}, (d) \textsf{H45} and (e) \textsf{DMC} tests.}
    \label{fig:meshes}
\end{figure}

To reduce the computational cost, the portions of the domain where the boundary conditions are applied or where the development of a crack is expected are discretized with a finer mesh.
As in the experimental results \cite{dmc_calibration1,dmc_calibration2,dmc_slides}, in all numerical tests a single crack starts from the tip of the notch and propagates upwards until complete failure.
Accordingly, we adopt the meshes illustrated in Fig.~\ref{fig:meshes} whose details in terms of number of elements and DOFs are summarized in Tab.~\ref{tab:dofsmesh}.
The notches are explicitly considered in the meshed geometries rather than modeled as a pre-crack using the phase-field variable, since they are introduced during manufacturing with a non-vanishing width.
\begin{table}[!bt]
    \caption{Mesh details for the TPB tests.}
    \label{tab:dofsmesh}
    \centering
    \footnotesize
    \begin{tabular}{P{2cm}P{2cm}P{2cm}P{2cm}P{2cm}P{2cm}}
        \toprule
        & \textbf{\textsf{HC} test} & \textbf{\textsf{HB} test} & \textbf{\textsf{HA} test} & \textbf{\textsf{H45}} & \textbf{\textsf{DMC} 
 test}\\
        \toprule
        \textbf{\# ELEMENTS} &125$\,$392 &215$\,$664 &354$\,$852 &415$\,$280 & 820$\,$063\\
        \textbf{\# DOFs}     &550$\,$036 &931$\,$776 &1$\,$515$\,$920 &1$\,$798$\,$724 &3$\,$425$\,$740\\
        \toprule
    \end{tabular}
\end{table}
Within the plane orthogonal to the notch, the \textsf{HC}, \textsf{HB} and \textsf{HA} meshes feature an average size of $h \simeq \ell/5$ = 0.125~mm and $h\simeq$ 1~mm in the fine and the coarse portions, respectively.
In the thickness direction, the size of the elements is larger and equal to about 0.397~mm in the fine and 3.175~mm in the coarse mesh regions (Figs.~\ref{fig:meshHC}-\ref{fig:meshHA})\footnote{For some representative cases, the results obtained with the meshes in Figs.~\ref{fig:meshHC}-\ref{fig:meshHA} are compared to those obtained with a mesh having the same element size of 0.125~mm (fine) and 1~mm (coarse) also in the thickness direction. The discrepancy between the two load-displacement curves lies within the 1~\% range, which is deemed acceptable considering the scatter of the experimental results.}.
In the \textsf{H45} and \textsf{DMC} meshes, we adopt a uniform size of the elements in all directions equal to 0.125~mm and 1~mm in the fine and coarse portions, respectively (Figs.~\ref{fig:meshH45}-\ref{fig:meshDMC}).
As visible in the details of the notches in Fig.~\ref{fig:meshes}, the refinement of the mesh is non-conforming, leading to the presence of hanging nodes which are automatically dealt with by the used \texttt{deal.ii} library by means of affine constraints \cite{dealII93,dealii2019design}.

\begin{figure}[!hbt]
    \centering
    \subfloat[]{
        \hspace{0.1cm}
        \includegraphics{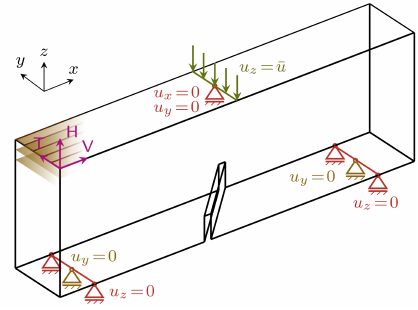}
        \hspace{0.1cm}
        \label{fig:disp_bc}
    }
    \subfloat[]{
        \hspace{0.1cm}
        \includegraphics{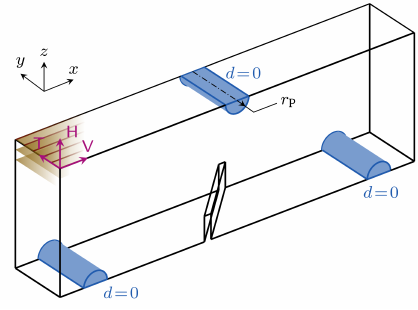}
        \hspace{0.1cm}
        \label{fig:pf_bc}
    }
    \caption{(a) Displacement boundary conditions and (b) phase-field internal conditions illustrated exemplarily for the \textsf{H45} specimen.}
    \label{fig:bcs}
\end{figure}

The applied boundary and internal conditions in terms of displacements and phase-field variable are illustrated  in Fig.~\ref{fig:bcs} on the representative example of the \textsf{H45} specimen.
The displacement boundary conditions are applied at the nominal contact line between loading or support rods and the specimen (Fig.~\ref{fig:disp_bc}).
For the phase-field variable, we apply a homogeneous Dirichlet condition within half-cylindrical sub-volumes with radius $r_{\text{P}}$ centered at the lines where the displacement boundary conditions are applied (Fig.~\ref{fig:pf_bc}).
This is done to prevent a spurious damage evolution in these portions of the domain, which are affected by highly localized reaction forces.
By means of preliminary analyses conducted by trial-and-error, we choose $r_{\text{P}}=$ 2.5~mm for specimens \textsf{HC}, \textsf{HB}, and \textsf{H45}; for specimens \textsf{HA} and \textsf{DMC}, since here the notch is farther away from midspan and a higher peak force is expected, we choose the slightly larger value $r_{\text{P}}=$ 3~mm.

We adopt a staggered solution procedure, which involves the alternate solution through a Newton-Raphson approach of the equilibrium equation \eqref{eq:equilibrium} and of the phase-field evolution equations \eqref{eq:phasefield} until the norms of the residuals fall below a given tolerance \cite{bourdin_numerical_2000} (set to $10^{-6}$ for the Newton-Raphson scheme and $10^{-3}$ for the staggered loop).
To enforce the irreversibility condition for the phase field \eqref{eq:phasefield}, although different alternatives are available \cite{gerasimov_penalization_2019}, we adopt the approach proposed in \cite{miehe_phase_2010}. We perform all computations with an in-house code based on the \texttt{C++} library \texttt{deal.ii} version 9.3.0 \cite{dealII93,dealii2019design}, on the \textit{Euler} High Performance Cluster of ETH Z\"urich using one single node with 128 CPU cores.

\section{Model calibration}\label{sec:parameter_calibration}
This section illustrates the procedure adopted to calibrate the model parameters summarized in Tab.~\ref{tab:probl}.
A total of eleven material constants are needed, including nine elastic parameters to determine the orthotropic stiffness tensor $\mathbb{C}$, the length-scale parameter $\ell$ and the fracture toughness $\mathcal{G}_{\text{c}}$, which we assume to be isotropic, as explained in Section~\ref{sct:model}.
However, as discussed in Section~\ref{sct:num_asp}, $\ell$ is considered here as a numerical parameter and its value is set to $\ell=$ 0.625~mm, which leaves us with ten remaining unknown parameters.
Also, a suitable dissipation function $w(d)$ must be selected.
Our basis for calibration are the results of the unconfined compressive, notched TPB and ultrasonic wave propagation tests available in \cite{dmc_calibration1,dmc_calibration2,dmc_slides}.
For the unconfined compressive and notched TPB tests we specifically consider the load $F$ vs. deflection $\bar{u}$ curves (see boundary conditions in Fig.~\ref{fig:disp_bc}).

\subsection{Preliminaries and calibration procedure outline}\label{sct:outline}
\begin{figure}[!hbt]
    \centering
    \includegraphics{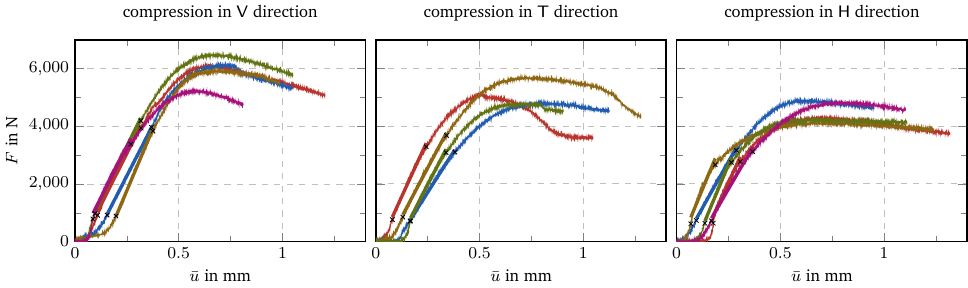}
    \label{fig:forcedisplacement_ucs}
    \caption{Unconfined compressive test results along the \Vdir, \Tdir and \Hdir directions, where $\bar{u}$ denotes the displacement and $F$ the reaction force.}
    \label{fig:compressive}
\end{figure}

\begin{figure}[!hbt]
    \centering
    \includegraphics{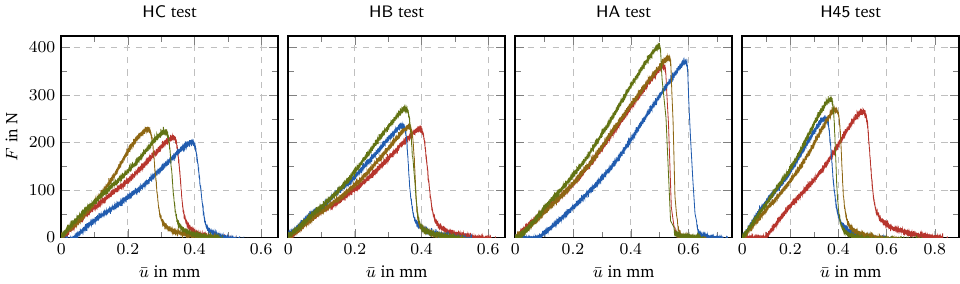}
    \label{fig:forcedisplacement_tpb}
    \caption{TPB test results for the  \textsf{HC}, \textsf{HB}, \textsf{HA} and \textsf{H45} specimens.}
    \label{fig:notchedTPB}
\end{figure}

The load vs. displacement curves of the unconfined compressive (Fig.~\ref{fig:compressive}) and notched TPB tests (Fig.~\ref{fig:notchedTPB}) display an initial linear elastic branch. For this reason, we select for the dissipation function the \textsf{AT1} model, see also Section~\ref{sct:model}.
Note that in Fig.~\ref{fig:notchedTPB} the curves for the \textsf{HA} specimens exhibit a higher peak force, confirming the choice of a larger $r_{\text{P}}$ for this geometry.

The calibration of the ten unknown parameters, i.e. the elastic parameters in $\mathbb{C}$ and the fracture toughness $\mathcal{G}_{\text{c}}$, is subdivided in two stages.
First, we use the unconfined compressive and the dynamic ultrasonic test results to obtain a first approximate estimate of the elastic tensor, denoted as $\mathbb{\tilde C}$.
As better explained in the following, this estimate cannot be considered representative of the behavior of the \textsf{DMC} notched TPB specimen, however we consider the ratios between the elastic parameters to be reliable.
We normalize $\tilde{C}_{ijkl}$ with $\tilde{C}_{\textsf{VVVV}}$ and assume
\begin{equation}
    \frac{C_{ijkl}}{C_{\textsf{VVVV}}}=\frac{\tilde{C}_{ijkl}}{\tilde{C}_{\textsf{VVVV}}}\quad \text{with}\quad i,j,k,l=\Vdir,\Tdir,\Hdir
    \label{eq:normalC}
\end{equation}
which reduces the number of unknown parameters to two, namely $C_{\textsf{VVVV}}$ and $\mathcal{G}_{\text{c}}$.
In the second stage of the calibration, we optimize these remaining quantities by means of an iterative optimization procedure based on the comparison between numerical and experimental load-deflection curves from the \textsf{HC}, \textsf{HB}, \textsf{HA} and \textsf{H45} notched TPB tests.

\subsection{First estimate of the elastic parameters}\label{sct:el_est}
We start by evaluating the main diagonal elements of $\tilde{\mathbb{C}}$ using the plane wave velocity measurements provided in \cite{dmc_calibration1,dmc_calibration2,dmc_slides}, where ultrasonic waves are let to propagate in cubic samples along the directions $\Vdir,\Tdir,\Hdir$ and receivers are placed so as to measure both longitudinal and shear wave velocities.
For a monochromatic plane wave propagating in a homogeneous and infinite anisotropic elastic medium, the relation between wave velocity and elastic parameters is obtained combining the plane wave equation and the equation of motion, giving {Christoffel}'s equation \cite{Baum1981}.
Using Einstein's summation convention, the latter reads \cite{fedorov_1968_theory}
\begin{equation}
    (\tilde{C}_{ijkl} w_j w_k - \rho s^2 \delta_{il}) \hat{u}_l = 0\,\quad \text{for}\quad i,j,k,l=\Vdir,\Tdir,\Hdir
    \label{eq:christoffel}
\end{equation}
where the vector $\bs{w}$ denotes the wave propagation direction and the vector $\hat{\bs{u}}$ defines the direction of particle motion.
We assume here that both coincide with the principal material directions \Vdir, \Tdir, \Hdir, meaning that they are either parallel for compressional waves or orthogonal to each other for shear waves.
Further, $\delta_{ij}$ is the Kronecker delta and $\rho =$ 1350~kg$\,$m$^{-3}$ (with a standard deviation of 17~kg$\,$m$^{-3}$) is the density of the material obtained as the average of three measurements on cubic samples \cite{dmc_slides}.
For notational convenience, we use the abbreviation $s_{mn}$ in the following to represent the wave velocity $s$ with propagation direction $m$ and particle movement direction $n$.
Hence, $s_{mn}$ represents the velocity of a longitudinal wave if $m=n$ and the velocity of a shear wave if $m\neq n$.
The average values of the wave velocities obtained from \cite{dmc_calibration1,dmc_calibration2,dmc_slides} are summarized in Tab.~\ref{tab:wave_vel} along with their standard deviations.

\begin{table}[b]
    \caption{Longitudinal and shear wave velocities: average values and standard deviations \cite{dmc_calibration1,dmc_calibration2,dmc_slides}. The first index denotes the direction of wave propagation and the second index is the direction of particle movement.}
    \label{tab:wave_vel}
    \centering
    \footnotesize
    \def\arraystretch{1.25}
    \begin{tabular}{P{5cm}P{2.5cm}P{2.5cm}}
        \toprule
        longitudinal wave velocities in m$/$s & \multicolumn{2}{c}{shear wave velocities in m$/$s}                     \\
        \toprule
        $s_{\textsf{VV}}= 2730$          & $s_{\textsf{VH}} = 1476$                      & $s_{\textsf{VT}} = 2020$ \\
        (st. dev. $= 12$)           & (st. dev. $= 63$)                        & (st. dev. $= 29$)   \\
        $s_{\textsf{TT}} = 2530$         & $s_{\textsf{TH}} = 1696$                      & $s_{\textsf{TV}} = 1980$ \\
        (st. dev. $= 83$)           & (st. dev. $= 71$)                        & (st. dev. $= 40$)   \\
        $s_{\textsf{HH}} = 2332$         & $s_{\textsf{HT}} = 1745$                      & $s_{\textsf{HV}} = 1799$ \\
        (st. dev. $= 84$)           & (st. dev. $= 62$)                        & (st. dev. $= 59$)   \\
        \toprule
    \end{tabular}
\end{table}

Eq. \eqref{eq:christoffel} with the measurements in Tab.~\ref{tab:wave_vel} gives \cite{Baum1981}
\begin{equation}
    \begin{aligned}
        \tilde{C}_{\textsf{VVVV}} & = \rho (s_{\textsf{VV}})^2 = {10061}\ \text{MPa}, \vphantom{\left( \frac{s_{22} + s_{33}}{2} \right)^2} \\
        \tilde{C}_{\textsf{TTTT}} & = \rho (s_{\textsf{TT}})^2 = {8641}\ \text{MPa}, \vphantom{\left( \frac{s_{22} + s_{33}}{2} \right)^2}  \\
        \tilde{C}_{\textsf{HHHH}} & = \rho (s_{\textsf{HH}})^2 = {7342}\ \text{MPa}, \vphantom{\left( \frac{s_{22} + s_{33}}{2} \right)^2}  \\
    \end{aligned}
    \qquad\qquad
    \begin{aligned}
        \tilde{C}_{\textsf{VTTV}} = \tilde{C}_{\textsf{TVVT}} & = \rho \left( \frac{s_{\textsf{VT}} + s_{\textsf{TV}}}{2} \right)^2 = {5400}\ \text{MPa ,} \\
        \tilde{C}_{\textsf{THHT}} = \tilde{C}_{\textsf{HTTH}} & = \rho \left( \frac{s_{\textsf{TH}} + s_{\textsf{HT}}}{2} \right)^2 = {3996}\ \text{MPa ,} \\
        \tilde{C}_{\textsf{HVVH}} = \tilde{C}_{\textsf{VHHV}} & = \rho \left( \frac{s_{\textsf{VH}} + s_{\textsf{HV}}}{2} \right)^2 = {3620}\ \text{MPa .} \\
    \end{aligned}\label{eq:est_c}
\end{equation}
Concerning the shear components in \eqref{eq:est_c},  the minor symmetry property $\tilde{C}_{ijkl}=\tilde{C}_{jilk}$ leads to $s_{jl}=s_{lj}$ but two separate measurements are available, which is why we use the mean value $(s_{jl}+s_{lj})/2$ for $j\neq l$.
For the off-diagonal components of $\tilde{\mathbb{C}}$, wave velocity measurements not along the principal directions and more involved calculations would be necessary, which is why the remaining three components of $\tilde{\mathbb{C}}$ are obtained following the standard relationships for orthotropic materials \cite{altenbach_2018_mechanics}
\begin{equation}
    \begin{aligned}
        \tilde{C}_{\textsf{VVTT}} & = \frac{\tilde{\nu}_{\textsf{TV}} + \tilde{\nu}_{\textsf{HV}}\tilde{\nu}_{\textsf{TH}}}{\xi}\tilde{E}_\Vdir ={8439}\ \text{MPa ,} \\
        \tilde{C}_{\textsf{TTHH}} & = \frac{\tilde{\nu}_{\textsf{HT}} + \tilde{\nu}_{\textsf{HV}}\tilde{\nu}_{\textsf{VT}}}{\xi}\tilde{E}_\Tdir = {6887}\ \text{MPa ,} \\
        \tilde{C}_{\textsf{HHVV}} & = \frac{\tilde{\nu}_{\textsf{HV}} + \tilde{\nu}_{\textsf{TV}}\nu_{\textsf{HT}}}{\xi}\tilde{E}_\Vdir= {7513}\ \text{MPa ,}          \\
    \end{aligned}
    \quad
    \begin{aligned}
        \\
        \text{with} \quad  \xi = 1 - \tilde{\nu}_{\textsf{VT}}\tilde{\nu}_{\textsf{TV}} - \tilde{\nu}_{\textsf{TH}}\tilde{\nu}_{\textsf{HT}} - \tilde{\nu}_{\textsf{HV}}\tilde{\nu}_{\textsf{VH}} - 2 \tilde{\nu}_{\textsf{VT}}\tilde{\nu}_{\textsf{TH}}\tilde{\nu}_{\textsf{HV}} \text{ .} \\
        \\
    \end{aligned}\label{eq:off_diag}
\end{equation}
It remains to explain how we obtain the Young's moduli $\tilde{E}_i$ and the Poisson's ratios $\tilde{\nu}_{ij}$ in \eqref{eq:off_diag}, with $i,j=\Vdir,\Tdir,\Hdir$. The elastic moduli are determined from unconfined compressive tests on cylindrical specimens with nominal length $l=$ 50.8~mm and radius $R=$ 12.7~mm. We first measure the uniaxial stiffnesses $\tilde{K}_i$ in the $\Vdir$, $\Tdir$ and $\Hdir$ directions as the slopes of the experimental $F$-$\bar{u}$ curves obtained by testing specimens aligned with the corresponding directions (Fig.~\ref{fig:compressive}), by means of a linear regression within the linear elastic regime, taken as the range from 15~\% to 65~\% of the peak force value $F_{\text{max}}$ (indicated with cross symbols in Fig.~\ref{fig:compressive}). Then, we compute the elastic moduli as \cite{dmc_calibration1,dmc_calibration2,dmc_slides}
\begin{equation}
\tilde{E}_i=\frac{\tilde{K}_i l}{\pi R^2} \quad \text{with}\quad i=\Vdir,\Tdir,\Hdir \text{ .}
    \label{eq:Young_mod}
\end{equation}
The obtained values of $\tilde{E}_i$ are summarized in Tab.~\ref{tab:YoungsPoisson}.

\begin{table}[!hbt]
    \caption{First estimate of Young's moduli and Poisson's ratios.}
    \label{tab:YoungsPoisson}
    \centering
    \footnotesize
    \begin{tabular}{P{2.5cm}P{2.5cm}P{2.5cm}P{1.5cm}P{1.5cm}P{1.5cm}}
        \toprule
        $\tilde{E}_{\textsf{V}}$ & $\tilde{E}_{\textsf{T}}$ & $\tilde{E}_{\textsf{H}}$ & $\tilde{\nu}_{\textsf{VT}}^{(\star)}$ & $\tilde{\nu}_{\textsf{TH}}^{(\star)}$ & $\tilde{\nu}_{\textsf{VH}}^{(\star)}$ \\[.5em]
        \toprule
        $1485$ MPa &$1365$ MPa &$1512$ MPa &$0.638$ &$0.338$ &$0.425$\\
        \toprule
        \multicolumn{6}{l}{\footnotesize $^{(\star)}$ Given $\nu_{ij}$, the value of $\nu_{ji}$ is obtained using the relationships \eqref{eq:ortho_id}.}
    \end{tabular}
\end{table}

The Poisson's ratios in \eqref{eq:off_diag} are obtained using the following equations
\begin{equation}
    \begin{aligned}
        \tilde{C}_{\textsf{VVVV}} & = \frac{1 - \tilde{\nu}_{\textsf{TH}}\tilde{\nu}_{\textsf{HT}}}{\xi}\tilde{E}_\Vdir\\
        \tilde{C}_{\textsf{TTTT}} & = \frac{1 - \tilde{\nu}_{\textsf{HV}}\tilde{\nu}_{\textsf{VH}}}{\xi}\tilde{E}_\Tdir\\
        \tilde{C}_{\textsf{HHHH}} & = \frac{1 - \tilde{\nu}_{\textsf{VT}}\tilde{\nu}_{\textsf{TV}}}{\xi}\tilde{E}_\Hdir\\
    \end{aligned}
\end{equation}
along with \cite{altenbach_2018_mechanics}
\begin{equation}
    \frac{\tilde{\nu}_{\textsf{VT}}}{\tilde{E}_\Vdir}
    = \frac{\tilde{\nu}_{\textsf{TV}}}{\tilde{E}_\Tdir} \quad \text{ , } \quad \frac{\tilde{\nu}_{\textsf{VH}}}{\tilde{E}_\Vdir}
    = \frac{\tilde{\nu}_{\textsf{HV}}}{\tilde{E}_\Hdir} \quad \text{ , } \quad \frac{\tilde{\nu}_{\textsf{TH}}}{\tilde{E}_\Tdir}
    = \frac{\tilde{\nu}_{\textsf{HT}}}{\tilde{E}_\Hdir}
    \label{eq:ortho_id}
\end{equation}
and the obtained values are summarized in Tab.~\ref{tab:YoungsPoisson}.

As mentioned in Section~\ref{sct:outline}, the estimates of the elastic parameters in \eqref{eq:est_c} and \eqref{eq:off_diag} cannot be directly adopted as final calibrated values due to different reasons. First, the estimates in \eqref{eq:est_c} are based on Christoffel's equation \eqref{eq:christoffel}, thus they should be intended as dynamic elastic parameters, and it is known that the quasi-static elastic parameters are usually lower than their dynamic counterparts by a factor that, for gypsum, can be as low as $1/4$ \cite{Vidales2021}.
Also, Christoffel's equation assumes that the tested material is homogeneous, while the investigated material is heterogeneous across different scales and includes interfacial zones (Section~\ref{sec:DMC_task}).
Finally, the experimental tests adopted to estimate the elastic parameters in \eqref{eq:off_diag} are representative of compressive rather than tensile stress states.
While a symmetric behavior under tension and compression may be expected for a homogeneous material, this assumption may easily break down for 
a heterogeneous material with interfacial zones.

Combining \eqref{eq:est_c} and \eqref{eq:off_diag} with the assumption \eqref{eq:normalC}, the elastic stiffness tensor can be reduced to
\begin{equation}\label{eq:C_ratio}
    \left[ \mathbb{C} \right] = C_{\textsf{VVVV}}
    \begin{bmatrix}
        1      & 0.8388 & 0.7467 & 0      & 0      & 0      \\
        0.8388 & 0.8588 & 0.6845 & 0      & 0      & 0      \\
        0.7467 & 0.6845 & 0.7297 & 0      & 0      & 0      \\
        0      & 0      & 0      & 0.3972 & 0      & 0      \\
        0      & 0      & 0      & 0      & 0.3598 & 0      \\
        0      & 0      & 0      & 0      & 0      & 0.5367 \\
    \end{bmatrix} \text{ .}
\end{equation}

\subsection{Parameters optimization procedure}\label{sct:optimization}
We now focus on the calibration of the two remaining parameters, $C_{\textsf{VVVV}}$ and $\mathcal{G}_{\text{c}}$. As mentioned earlier, this is based on an optimization procedure which aims at minimizing the discrepancies between the load-deflection curves for the  TPB characterization specimens obtained numerically and experimentally, denoted respectively as $F_{\text{num},\,i}(\bar{u}\,|\,C_{\textsf{VVVV}},\,\mathcal{G}_{\text{c}})$ and $F_{\text{exp},\,i}(\bar{u})$ with $i=$ \textsf{HC}, \textsf{HB}, \textsf{HA}, \textsf{H45}.

\begin{figure}[!hbt]
    \centering
    \subfloat[]{\label{fig:forcedisplacement_averaging}
        \raisebox{0.325cm}{\hspace*{-0.15cm}\includegraphics{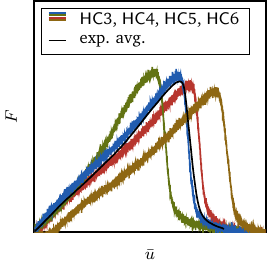}}
    }
    \subfloat[]{\label{fig:aerial_difference}
        \raisebox{0.325cm}{\hspace*{-0.15cm}\includegraphics{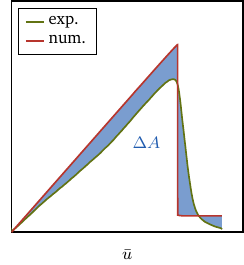}}
    }
    \hspace*{0.5cm}
    \subfloat[]{\label{fig:minimization_problem}
        \includegraphics{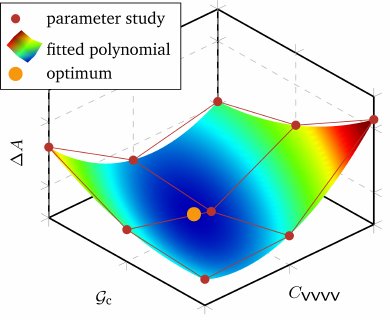}
    }
    \caption{(a) Illustrative scheme of experimental and average load-deflection curves $F\sdash\bar u$ (on the example of the \textsf{HC} specimen). (b) Cost function for the optimization procedure. (c) Fitting of a second-order polynomial approximation of the cost function using a sampling grid of nine points and definition of the optimum. }
    \label{eq:calibration_problem}
\end{figure}

For given values of $C_{\textsf{VVVV}}$ and $\mathcal{G}_{\text{c}}$, the numerical curves are obtained by predicting the behavior until failure of the  TPB characterization specimens, using the model and numerical setup illustrated in Sections~\ref{sct:model}-\ref{sct:num_asp} along with \eqref{eq:C_ratio}.
Each experimental curve for a TPB specimen is taken as the average of the curves obtained from all repetitions of the respective test.
To obtain these average curves, we first apply a moving average filter and then normalize the data points in each curve with the respective peak force and its associated displacement.
We then shift the curves such that the initial instant of the test is the point at which this normalized force becomes and remains positive for the first time.
These curves are then resampled using a common grid of normalized displacements by interpolating the normalized force between the experimental measurements of each individual test.
Since the normalized curves are now sampled with a common, equidistant grid of normalized displacements, the average normalized force across the different tests can be computed for each of the sampling points.
Finally, this average of the normalized and resampled curves is scaled up by the arithmetic mean of the individual normalization factors for the respective peak force and its associated displacement to obtain the average experimental curve (Fig.~\ref{fig:forcedisplacement_averaging}).

To setup the optimization problem, we use as cost function the integral over the loading history of the mismatch between experimental and numerical force-deflection curves (Fig.~\ref{fig:aerial_difference}), i.e.
\begin{equation}
    \Delta A_i\left(C_{\textsf{VVVV}},\,\mathcal{G}_{\text{c}}\right)=\int_{\text{hist.}} \! \left| F_{\text{num},\,i}(\bar{u}\,|\,C_{\textsf{VVVV}},\,\mathcal{G}_{\text{c}}) - F_{\text{exp},\,i}(\bar{u}) \right| \mathrm{d}\bar{u}\,,\quad \text{for}\quad i=\textsf{HC},\  \textsf{HB},\  \textsf{HA},\  \textsf{H45} \text{ .}
    \label{eq:error_fct}
\end{equation}

Clearly, the cost function $\Delta A_i\left(C_{\textsf{VVVV}},\,\mathcal{G}_{\text{c}}\right)$ cannot be obtained analytically, but it can be sampled at given values of $C_{\textsf{VVVV}}$ and $\mathcal{G}_{\text{c}}$.
Thus, we assume that \eqref{eq:error_fct} can be represented by a second-order polynomial of the form
\begin{equation}
\begin{split}
    \Delta A_i\left(C_{\textsf{VVVV}},\,\mathcal{G}_{\text{c}}\right) = p_{i,0} + p_{i,1} C_{\textsf{VVVV}} + p_{i,2} \mathcal{G}_{\text{c}} + p_{i,3} C_{\textsf{VVVV}}^2 + p_{i,4}& C_{\textsf{VVVV}} \mathcal{G}_{\text{c}} + p_{i,5} \mathcal{G}_{\text{c}}^2\,,\\ &\text{for}\quad i=\textsf{HC},\  \textsf{HB},\  \textsf{HA},\  \textsf{H45} \text{ ,}
    \end{split}
    \label{eq:polynomial_ansatz}
\end{equation}
and we obtain the coefficients $p_{i,k}$ by means of a least-square fitting, minimizing the mismatch between a grid of sampled values and its representation \eqref{eq:polynomial_ansatz} as illustrated in Fig.~\ref{fig:minimization_problem}.
From \eqref{eq:polynomial_ansatz}, the optimal values $\left(C_{\textsf{VVVV}}^*,\,\mathcal{G}_{\text{c}}^*\right)_i$can be obtained by solving analytically the minimization problem
\begin{equation}
    \left(C_{\textsf{VVVV}}^*,\,\mathcal{G}_{\text{c}}^*\right)_i= \underset{C_{\textsf{VVVV}},\mathcal{G}_{\text{c}}}{\arg \min}\ \Delta A_i\left(C_{\textsf{VVVV}},\,\mathcal{G}_{\text{c}}\right)\,,\quad \text{for}\quad i=\textsf{HC},\  \textsf{HB},\  \textsf{HA},\  \textsf{H45} \text{ .}
    \label{eq:min_prbl}
\end{equation}
To ensure the selection of the best set of parameters, the optimization procedure is performed iteratively by progressively restricting the range of parameters $\left(C_{\textsf{VVVV}},\,\mathcal{G}_{\text{c}}\right)$ used to sample the cost function \eqref{eq:error_fct}.
The procedure is initialized selecting a rather broad range for $\left(C_{\textsf{VVVV}},\,\mathcal{G}_{\text{c}}\right)$; then, at each iteration we refine the grid of sampling points in the neighborhood of the optimum set of parameters found at the previous iteration.
The process is terminated after three iterations of parameter grid refinement, whereby the first and second iterations use a grid of $3\times 3$, and the final iteration uses a grid of $5 \times 5$ sampled points in the $\left(C_{\textsf{VVVV}},\,\mathcal{G}_{\text{c}}\right)$ space.

Since the proposed procedure is performed separately for all TPB characterization tests, we obtain four sets of parameters summarized in Tab.~\ref{tab:parameters_opt}. Their average gives the final calibrated values $C_{\textsf{VVVV}}^* =$ 2057~MPa and $\mathcal{G}_{\text{c}}^* =$ 9.23 $\cdot 10^{-2}$ MPa mm.

\begin{table}[!hbt]
    \caption{Optimized parameters for the different TPB characterization tests.}
    \label{tab:parameters_opt}
    \centering
    \footnotesize
    \begin{tabular}{P{2cm}P{1.5cm}P{1.5cm}P{1.5cm}P{1.5cm}}
        \toprule
        & \textbf{\textsf{HC} test} & \textbf{\textsf{HB} test} & \textbf{\textsf{HA} test} & \textbf{\textsf{H45} test}\\[.5em]
        \toprule
        $\displaystyle C_{\textsf{VVVV}}^*$ in MPa &$2118$ &$1958$ &$2040$ &$2112$\\[.5em]
        $\displaystyle \mathcal{G}_{\text{c}}^*$ in MPa mm &$8.30\cdot10^{-2}$ &$8.62\cdot10^{-2}$ &$8.55\cdot10^{-2}$ &$11.45\cdot10^{-2}$\\[.2em]
        \toprule
    \end{tabular}
\end{table}

\subsection{Calibrated material parameters}\label{sct:summary}
\begin{figure}[!hbt]
    \centering
    \subfloat[]{\label{fig:orthotropy_youngsmod}
        \includegraphics{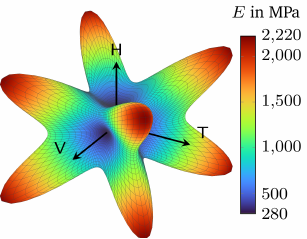}
    }
    \subfloat[]{\label{fig:orthotropy_poissonavg}
        \includegraphics{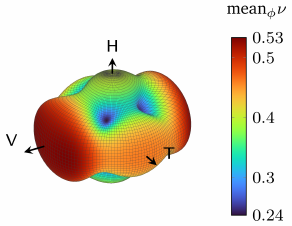}
    }
    \subfloat[]{\label{fig:orthotropy_poissonmax}
        \includegraphics{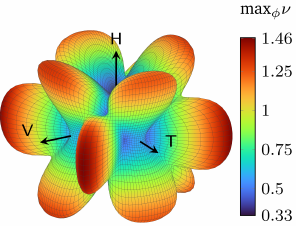}
    }
    \caption{Visualization of (a) the Young's modulus $E$ for different test directions $\bs{d}$ as well as (b) the arithmetic mean and (c) the maximum of the Poisson's ratio $\nu$ of different angles $\phi$ around it.}
    \label{eq:orthotropy}
\end{figure}

Following the procedure described in Sections.~\ref{sct:el_est} and \ref{sct:optimization}, we obtain
\begin{equation}
    \left[ \mathbb{C} \right] =
    \begin{bmatrix}
        2057 & 1725 & 1536 & 0      & 0      & 0      \\
        1725 & 1767 & 1408 & 0      & 0      & 0      \\
        1536 & 1408 & 1501 & 0      & 0      & 0      \\
        0    & 0    & 0    & 817    & 0      & 0      \\
        0    & 0    & 0    & 0      & 740    & 0      \\
        0    & 0    & 0    & 0      & 0      & 1104   \\
    \end{bmatrix}
    \text{MPa}
\end{equation}
for the orthotropic elastic stiffness tensor and
\begin{equation}
    \mathcal{G}_{\text{c}} = 9.23\cdot10^{-2} \text{ MPa mm}
\end{equation}
for the fracture toughness.
Note that the obtained values of the elastic parameters are about $1/5$ of those obtained for $\tilde{\mathbb{C}}$, which we explain with the model simplifications discussed in Section~\ref{sct:model} as well as the fact that the estimates are based on a combination of dynamic and quasi-static experiments.
A visualization of the anisotropy of the elastic properties is given in Fig. \ref{eq:orthotropy}, based on the procedure and implementation presented in \cite{Nordmann_2018_visualization}.
In particular, Fig. \ref{fig:orthotropy_youngsmod} shows the Young's modulus one would obtain when performing a uni-axial tensile test in the direction $\bs{d}$ (described in a spherical coordinate system), clearly indicating the three symmetry planes and that the strongest directions of the material are at an angle of roughly 45$^\circ$ to the manufacturing planes.
To obtain the Poisson's ratio for such uniaxial test, a normal vector giving the direction of lateral contraction is required, which is defined vertical to $\bs{d}$ and at an angle of $\phi$.
The averaged as well as the maximum value for the Poisson's ratio for this angle $\phi$ around the test direction $\bs{d}$ are plotted in Figs. \ref{fig:orthotropy_poissonavg} and \ref{fig:orthotropy_poissonmax}, respectively, again indicating the three symmetry planes for orthotropic material.

\section{Comparison between numerical and experimental results}\label{sec:results_and_discussion}
This section illustrates the comparison between experimental results and numerical predictions obtained using the calibrated parameters.
We discuss first the TPB characterization tests and then the blind prediction of the \textsf{DMC} test following the task summarized in Section~\ref{sec:DMC_task}.

\subsection{Comparison with the TPB characterization  test results}\label{sct:comp_calibr}
\begin{figure}[!hbt]
    \centering\includegraphics{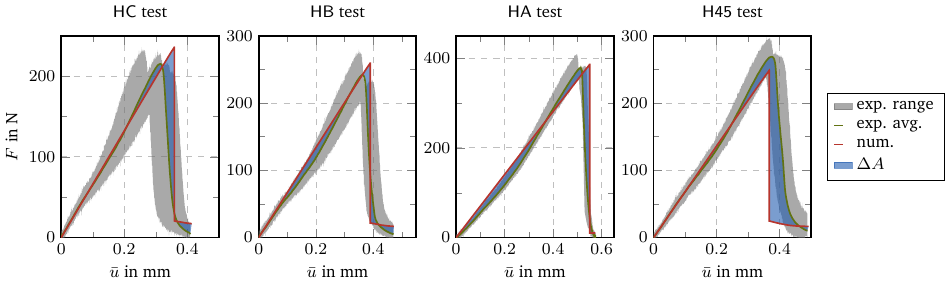}
    \caption{Comparison between numerical and experimental load-deflection curves for the \textsf{HC},  \textsf{HB}, \textsf{HA} and \textsf{H45} tests.}
    \label{fig:forcedisplacement_calibration}
\end{figure}

We start by comparing the experimental and numerical load-deflection curves for the \textsf{HC}, \textsf{HB}, \textsf{HA} and \textsf{H45} specimens (Fig.~\ref{fig:forcedisplacement_calibration}).
Considering the experimental scatter and that the numerical curves are all obtained with a unique set of material parameters, we consider the agreement very satisfactory.
In particular, the peak load and the initial elastic branch are well captured in all cases.
After the peak, the numerical curves display an abrupt load drop due to the assumption of a perfectly brittle material. The experimental post-peak behavior is slightly smoother, probably due to cohesive-frictional phenomena neglected in the model.
However, the overall agreement confirms that a brittle fracture model predicts the real behavior of the material reasonably well.

In the late stage of the tests, the numerical predictions present an almost horizontal plateau while the experimental curves show a slow but steady decrease of the load (Fig.~\ref{fig:forcedisplacement_calibration}).
This is mostly due to the homogeneous Dirichlet conditions imposed on the phase-field variable, which prevent the crack from reaching the upper surface of the specimen. This leads to a residual load bearing capacity that, however, amounts to less than 10~\% of the peak force.

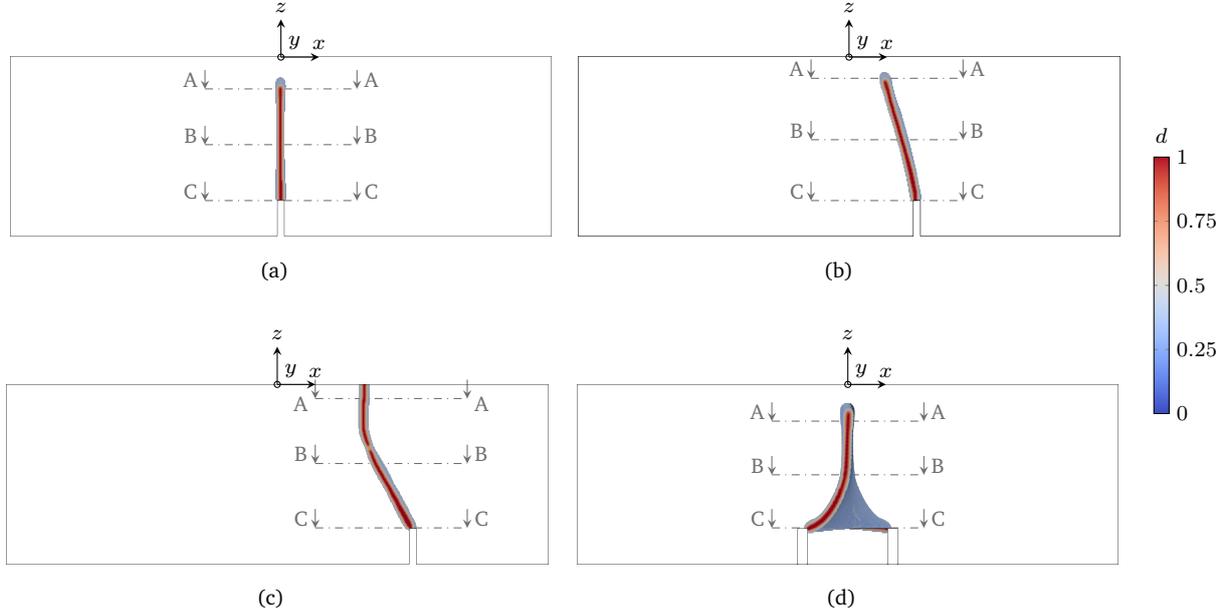
\begin{figure}[!hbt]
    \subfloat[]{\label{fig:crack_HC}
         \centering
         \begin{tikzpicture}
             \footnotesize
             \draw[grau, thin, dash dot] (-1,0.7658) -- (1,0.7658); 
             \draw[grau, thin, -stealth] (-1,0.7658+0.25) -- (-1,0.7658);
             \draw[grau, thin, -stealth] (1,0.7658+0.25) -- (1,0.7658);
             \node[grau, anchor=east] at (-1,0.7658+0.125) {A};
             \node[grau, anchor=west] at (1,0.7658+0.125) {A};
             \draw[grau, thin, dash dot] (-1,0.0263) -- (1,0.0263); 
             \draw[grau, thin, -stealth] (-1,0.0263+0.25) -- (-1,0.0263);
             \draw[grau, thin, -stealth] (1,0.0263+0.25) -- (1,0.0263);
             \node[grau, anchor=east] at (-1,0.0263+0.125) {B};
             \node[grau, anchor=west] at (1,0.0263+0.125) {B};
             \draw[grau, thin, dash dot] (-1,-0.715) -- (1,-0.715); 
             \draw[grau, thin, -stealth] (-1,-0.715+0.25) -- (-1,-0.715);
             \draw[grau, thin, -stealth] (1,-0.715+0.25) -- (1,-0.715);
             \node[grau, anchor=east] at (-1,-0.715+0.125) {C};
             \node[grau, anchor=west] at (1,-0.715+0.125) {C};
             \draw[thin, -stealth] (0,1.187) -- (0.5,1.187) node[anchor=south] at (0.5,1.187) {$x$};
             \draw[thin, -stealth] (0,1.187) -- (0,1.187+0.5) node[anchor=south] at (0,1.187+0.5) {$z$};
             \draw[thin] (0,1.187) circle (0.04) node[anchor=south west] at (0,1.187) {$y$};
             \node[anchor=center] at (0,0) {\includegraphics[width=7.125cm]{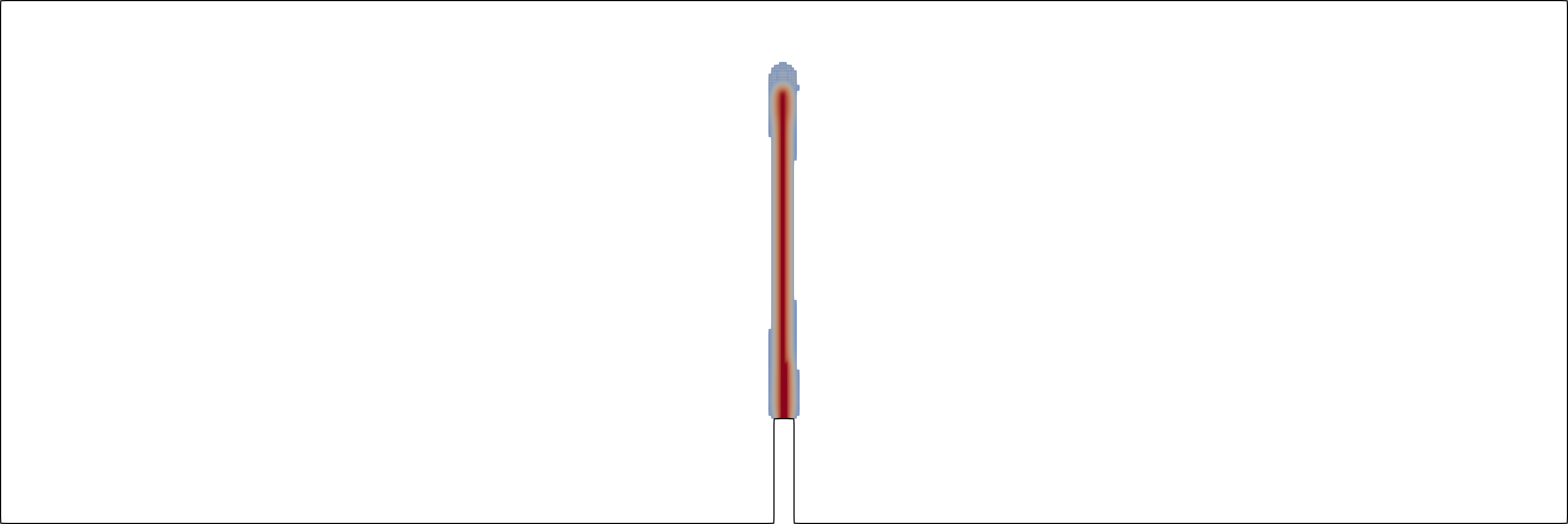}};
         \end{tikzpicture}
     }
    \hspace*{-0.18cm}
    \subfloat[]{ \label{fig:crack_HB}
         \centering
         \begin{tikzpicture}
             \footnotesize
             \draw[grau, thin, dash dot] (-0.5,0.9062) -- (1.5,0.9062); 
             \draw[grau, thin, -stealth] (-0.5,0.9062+0.25) -- (-0.5,0.9062);
             \draw[grau, thin, -stealth] (1.5,0.9062+0.25) -- (1.5,0.9062);
             \node[grau, anchor=east] at (-0.5,0.9062+0.125) {A};
             \node[grau, anchor=west] at (1.5,0.9062+0.125) {A};
             \draw[grau, thin, dash dot] (-0.5,0.0919) -- (1.5,0.0919); 
             \draw[grau, thin, -stealth] (-0.5,0.0919+0.25) -- (-0.5,0.0919);
             \draw[grau, thin, -stealth] (1.5,0.0919+0.25) -- (1.5,0.0919);
             \node[grau, anchor=east] at (-0.5,0.0919+0.125) {B};
             \node[grau, anchor=west] at (1.5,0.0919+0.125) {B};
             \draw[grau, thin, dash dot] (-0.5,-0.715) -- (1.5,-0.715); 
             \draw[grau, thin, -stealth] (-0.5,-0.715+0.25) -- (-0.5,-0.715);
             \draw[grau, thin, -stealth] (1.5,-0.715+0.25) -- (1.5,-0.715);
             \node[grau, anchor=east] at (-0.5,-0.715+0.125) {C};
             \node[grau, anchor=west] at (1.5,-0.715+0.125) {C};
             \draw[thin, -stealth] (0,1.187) -- (0.5,1.187) node[anchor=south] at (0.5,1.187) {$x$};
             \draw[thin, -stealth] (0,1.187) -- (0,1.187+0.5) node[anchor=south] at (0,1.187+0.5) {$z$};
             \draw[thin] (0,1.187) circle (0.04) node[anchor=south west] at (0,1.187) {$y$};
             \node[anchor=center] at (0,0) {\includegraphics[width=7.125cm]{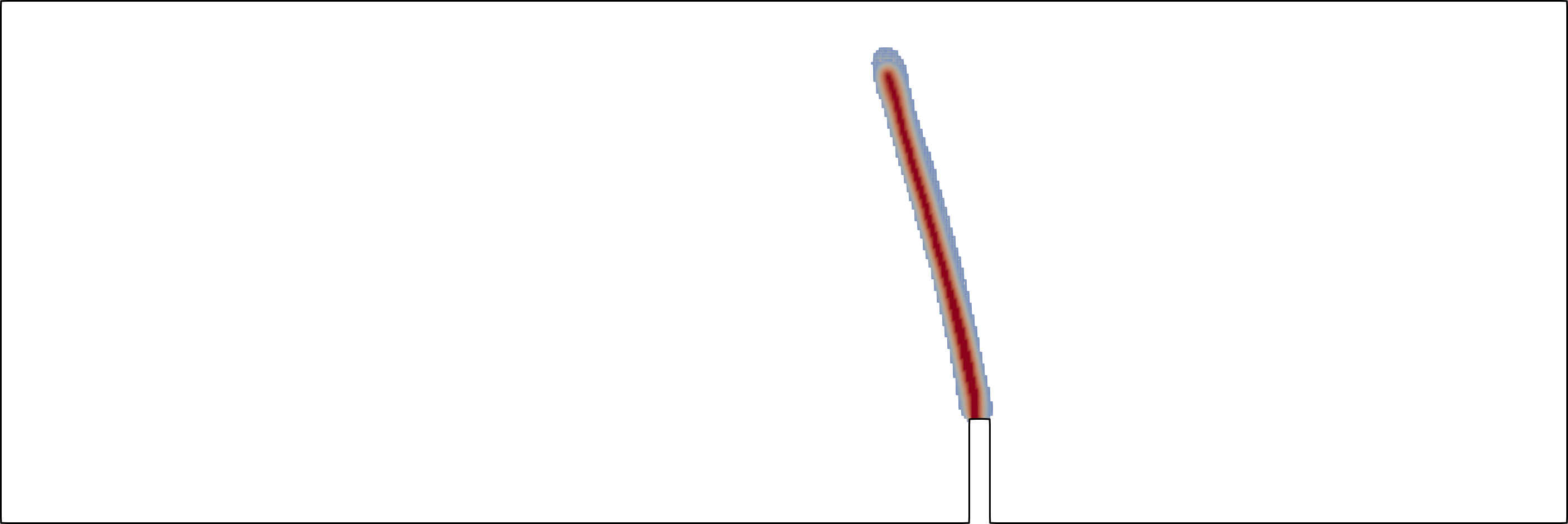}};
         \end{tikzpicture}}
    \hfill\\[-2.2cm]
    \subfloat[]{\label{fig:crack_HA}
         \centering
         \begin{tikzpicture}
             \footnotesize
             \draw[grau, thin, dash dot] (0.5,0.9998) -- (2.5,0.9998); 
             \draw[grau, thin, -stealth] (0.5,0.9998+0.25) -- (0.5,0.9998);
             \draw[grau, thin, -stealth] (2.5,0.9998+0.25) -- (2.5,0.9998);
             \node[grau, anchor=north east] at (0.5,0.9998+0.125) {A};
             \node[grau, anchor=north west] at (2.5,0.9998+0.125) {A};
             \draw[grau, thin, dash dot] (0.5,0.1387) -- (2.5,0.1387); 
             \draw[grau, thin, -stealth] (0.5,0.1387+0.25) -- (0.5,0.1387);
             \draw[grau, thin, -stealth] (2.5,0.1387+0.25) -- (2.5,0.1387);
             \node[grau, anchor=east] at (0.5,0.1387+0.125) {B};
             \node[grau, anchor=west] at (2.5,0.1387+0.125) {B};
             \draw[grau, thin, dash dot] (0.5,-0.715) -- (2.5,-0.715); 
             \draw[grau, thin, -stealth] (0.5,-0.715+0.25) -- (0.5,-0.715);
             \draw[grau, thin, -stealth] (2.5,-0.715+0.25) -- (2.5,-0.715);
             \node[grau, anchor=east] at (0.5,-0.715+0.125) {C};
             \node[grau, anchor=west] at (2.5,-0.715+0.125) {C};
             \draw[thin, -stealth] (0,1.187) -- (0.5,1.187) node[anchor=south] at (0.5,1.187) {$x$};
             \draw[thin, -stealth] (0,1.187) -- (0,1.187+0.5) node[anchor=south] at (0,1.187+0.5) {$z$};
             \draw[thin] (0,1.187) circle (0.04) node[anchor=south west] at (0,1.187) {$y$};
             \node[anchor=center] at (0,0) {\includegraphics[width=7.125cm]{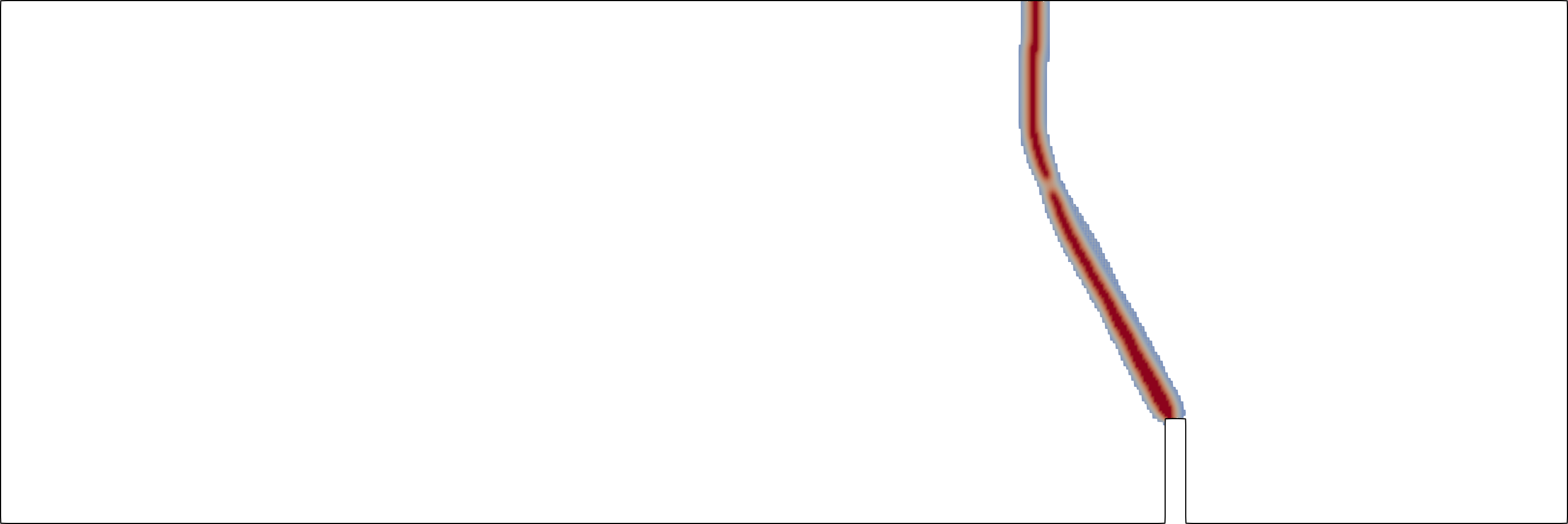}};
         \end{tikzpicture}}
    \vspace*{0cm}
    \subfloat[]{\label{fig:crack_H45}
         \centering
         \begin{tikzpicture}
             \footnotesize
             \draw[grau, thin, dash dot] (-1,0.7003) -- (1,0.7003); 
             \draw[grau, thin, -stealth] (-1,0.7003+0.25) -- (-1,0.7003);
             \draw[grau, thin, -stealth] (1,0.7003+0.25) -- (1,0.7003);
             \node[grau, anchor=east] at (-1,0.7003+0.125) {A};
             \node[grau, anchor=west] at (1,0.7003+0.125) {A};
             \draw[grau, thin, dash dot] (-1,-0.0111) -- (1,-0.0111); 
             \draw[grau, thin, -stealth] (-1,-0.0111+0.25) -- (-1,-0.0111);
             \draw[grau, thin, -stealth] (1,-0.0111+0.25) -- (1,-0.0111);
             \node[grau, anchor=east] at (-1,-0.0111+0.125) {B};
             \node[grau, anchor=west] at (1,-0.0111+0.125) {B};
             \draw[grau, thin, dash dot] (-1,-0.715) -- (1,-0.715); 
             \draw[grau, thin, -stealth] (-1,-0.715+0.25) -- (-1,-0.715);
             \draw[grau, thin, -stealth] (1,-0.715+0.25) -- (1,-0.715);
             \node[grau, anchor=east] at (-1,-0.715+0.125) {C};
             \node[grau, anchor=west] at (1,-0.715+0.125) {C};
             \draw[thin, -stealth] (0,1.187) -- (0.5,1.187) node[anchor=south] at (0.5,1.187) {$x$};
             \draw[thin, -stealth] (0,1.187) -- (0,1.187+0.5) node[anchor=south] at (0,1.187+0.5) {$z$};
             \draw[thin] (0,1.187) circle (0.04) node[anchor=south west] at (0,1.187) {$y$};
             \node[anchor=center] at (0,0) {\includegraphics[width=7.125cm]{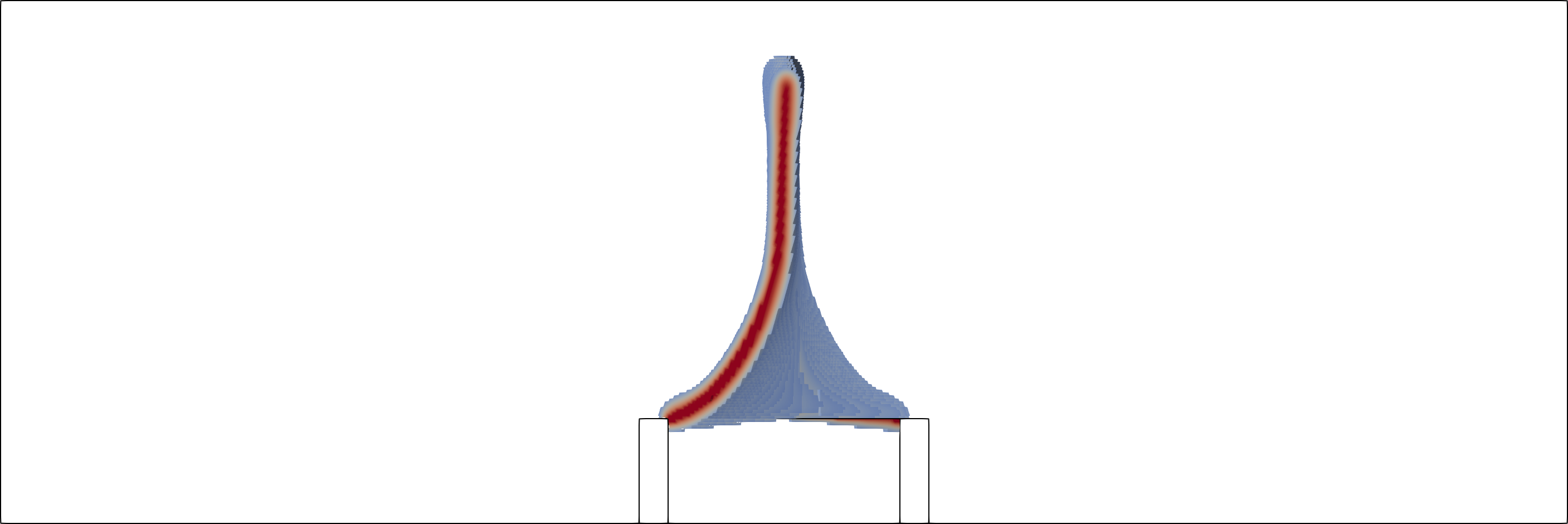}};
         \end{tikzpicture}}
     \begin{subfigure}[b]{0.05\linewidth}
         \hspace*{0.3cm}\raisebox{1.9cm}{\includegraphics{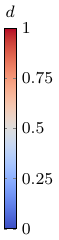}}
     \end{subfigure}
     \caption{Crack path at the last load step obtained numerically for the (a) \textsf{HC}, (b)  \textsf{HB}, (c) \textsf{HA} and (d) \textsf{H45} specimens in terms of the phase-field variable $d$. (All points with $d\le$ 0.25 are hidden for the sake of clarity).}
     \label{fig:crack_surfaces_images_thres}\setcounter{subfigure}{0}
 \end{figure}

The numerical crack patterns obtained at the last load step are compared in Fig.~\ref{fig:crack_surfaces_images_thres}, where the phase-field values $d\ge$ 0.25 are shown.
Here the geometry and loading conditions induce different types of fracture including pure mode I (Fig.~\ref{fig:crack_HC}), mixed mode I+II (Figs.~\ref{fig:crack_HB}-\ref{fig:crack_HA}) and mixed mode I+III (Fig.~\ref{fig:crack_H45}).
As expected, the crack starts at the notch tip and propagates upwards until reaching either the region with homogeneous Dirichlet conditions for the phase field (Figs~\ref{fig:crack_HC}, \ref{fig:crack_HB} and \ref{fig:crack_H45}) or the upper edge of the specimen (Fig.~\ref{fig:crack_HA}).
The initial crack evolution happens unstably at the peak load, with an abrupt propagation of the crack front along a large part of the beam height; very limited further propagation takes place after the peak.
This agrees with the digital image correlation measurements of the experiments \cite{dmc_calibration2}, where an abrupt propagation of the crack front can be observed.

As anticipated in Section~\ref{sct:num_asp}, in the  \textsf{HC}, \textsf{HB} and \textsf{HA} specimens the gradient of the phase-field variable is negligible in the thickness direction, therefore a coarser mesh can be adopted in that direction.
This is not the case for the \textsf{H45} specimen, where a twisting of the crack front takes place as it propagates from the notch to the upper edge of the specimen.

\begin{figure}[!hbt]
    \centering\includegraphics{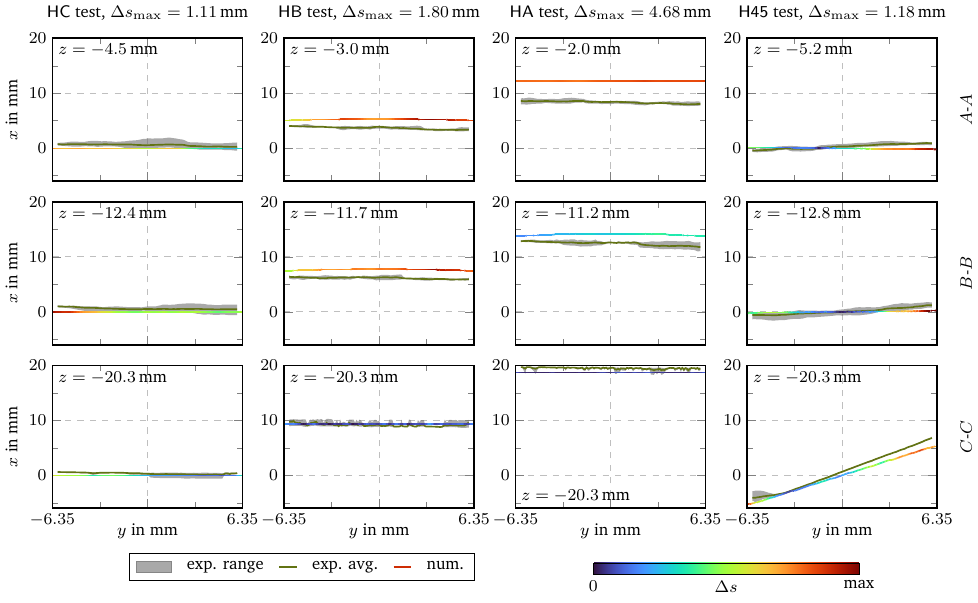}
    \caption{Comparison between numerical and experimental crack paths within the three horizontal sections $A\sdash A$, $B\sdash B$ and $C\sdash C$ illustrated in Fig.~\ref{fig:crack_surfaces_images_thres} for the \textsf{HC}, \textsf{HB}, \textsf{HA} and \textsf{H45} tests.}
    \label{fig:cracksurface_cuts_calibration}
\end{figure}

\begin{figure}[!hbt]
    \centering
    \subfloat[]{
    \begin{tabular}{ccc}
\multicolumn{3}{c}{\footnotesize \textsf{HC} test, $\Delta {s}_{\text{max}} =$ 1.11~mm} \\
        \fbox{\includegraphics[height=3.5cm]{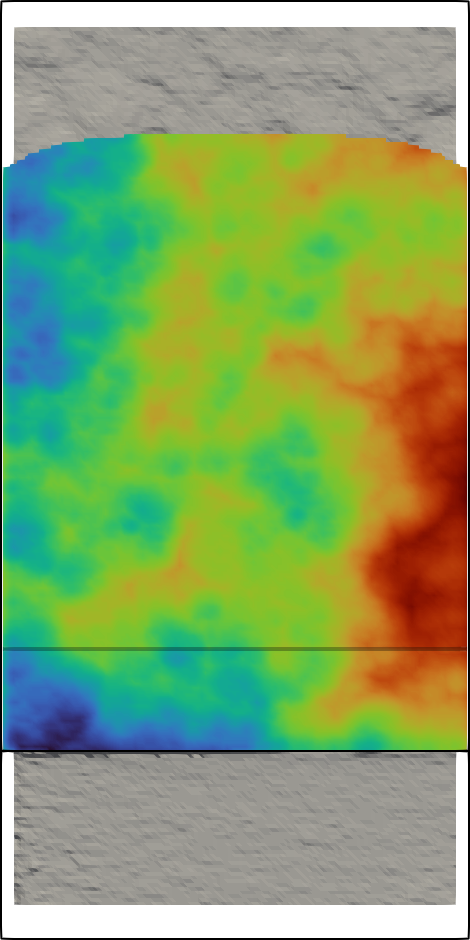}} & \fbox{\adjincludegraphics[height=3.5cm, trim={{.35\width} 0 {.35\width} 0}, clip]{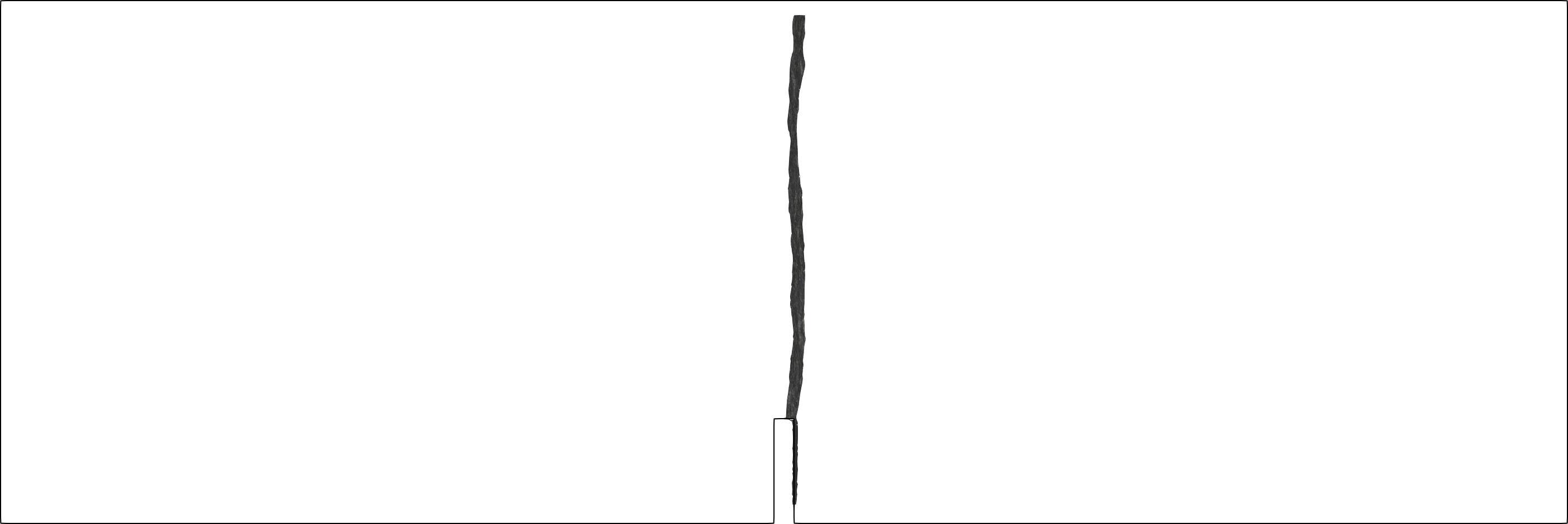}} & \fbox{\includegraphics[height=3.5cm]{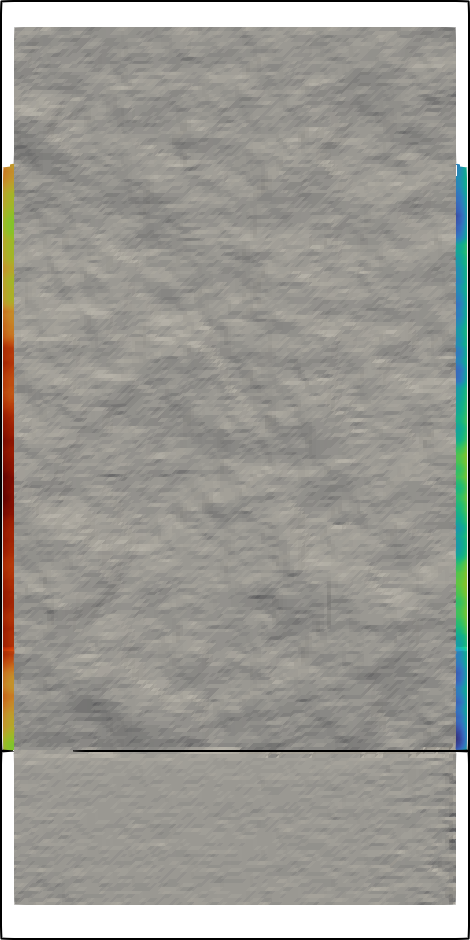}} \\
        {\footnotesize (left)} &{\footnotesize (front)} &{\footnotesize (right)}\\[-.5em]
    \end{tabular}}
    \subfloat[]{
    \begin{tabular}{ccc}
\multicolumn{3}{c}{\footnotesize \textsf{HB} test, $\Delta {s}_{\text{max}} =$ 1.80~mm} \\
        \fbox{\includegraphics[height=3.5cm]{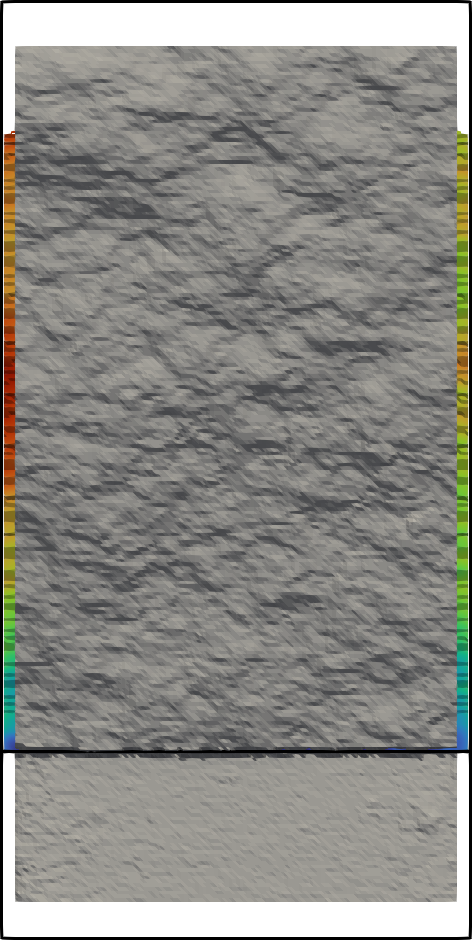}} & \fbox{\adjincludegraphics[height=3.5cm, trim={{.4\width} 0 {.3\width} 0}, clip]{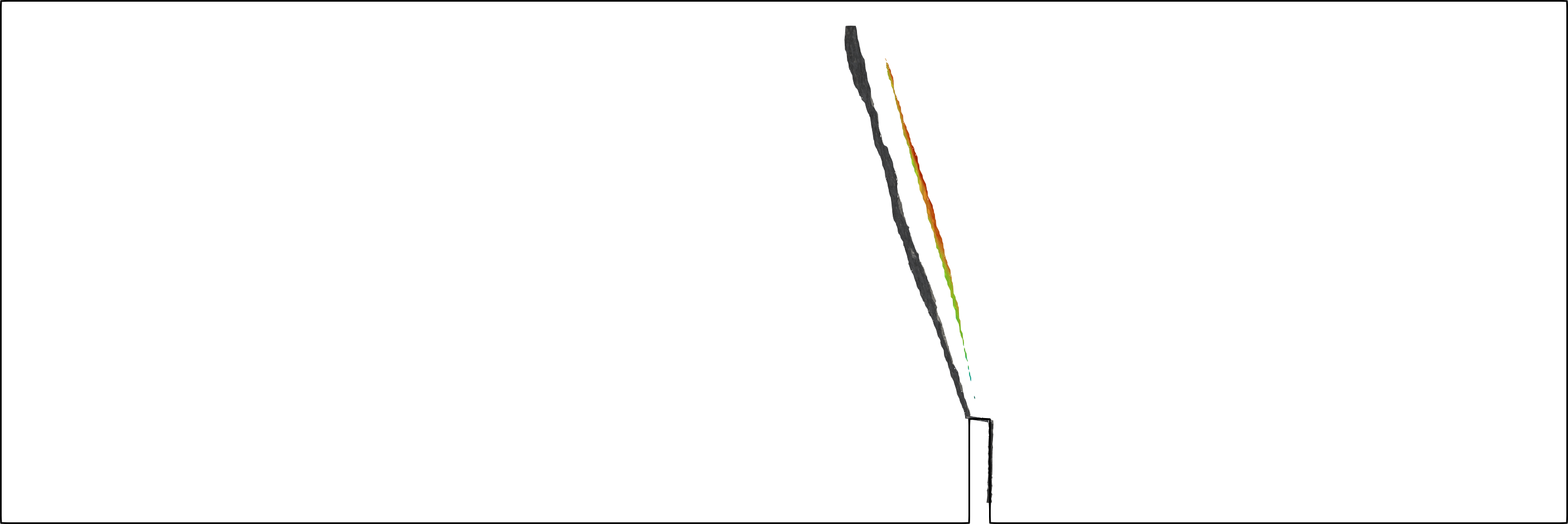}} & \fbox{\includegraphics[height=3.5cm]{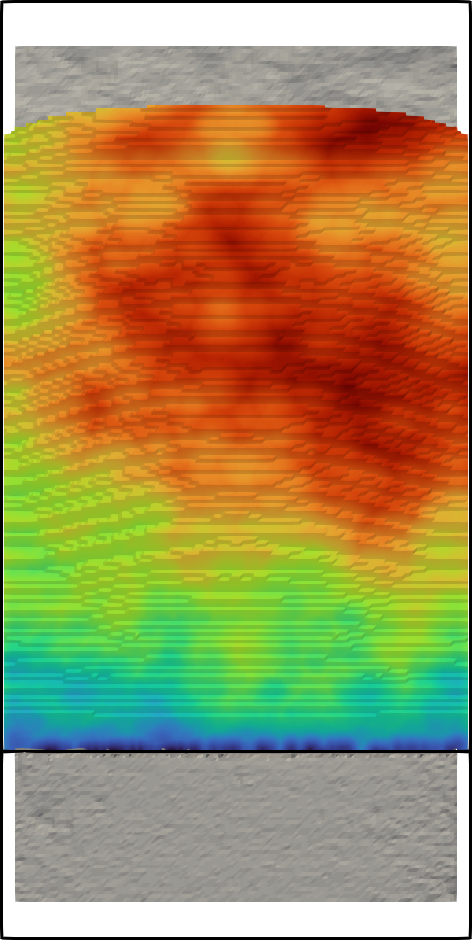}} \\
        {\footnotesize (left)} &{\footnotesize (front)} &{\footnotesize (right)}\\[-.5em]
    \end{tabular}}\\[0.2cm]
    \subfloat[]{
    \begin{tabular}{ccc}
\multicolumn{3}{c}{\footnotesize \textsf{HA} test, $\Delta {s}_{\text{max}} =$ 4.68~mm} \\
        \fbox{\includegraphics[height=3.5cm]{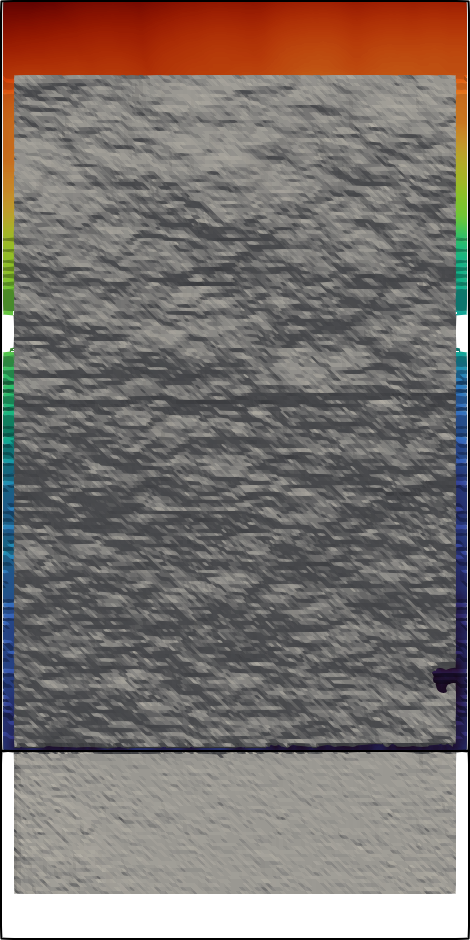}} & \fbox{\adjincludegraphics[height=3.5cm, trim={{.55\width} 0 {.15\width} 0}, clip]{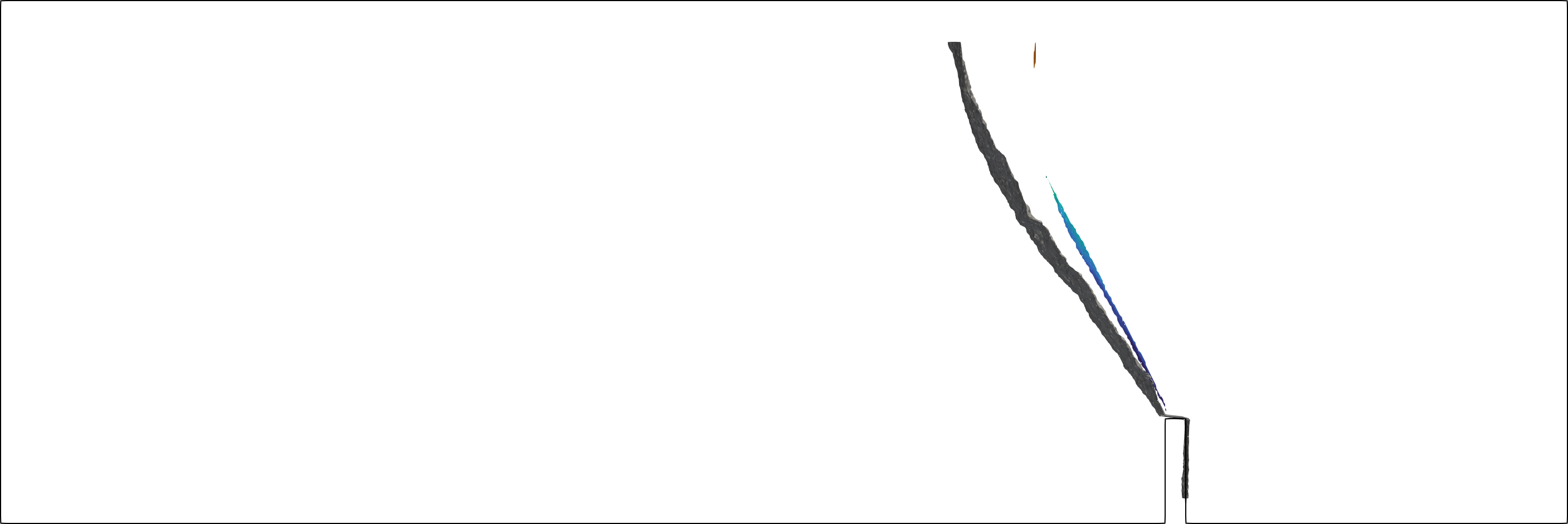}} & \fbox{\includegraphics[height=3.5cm]{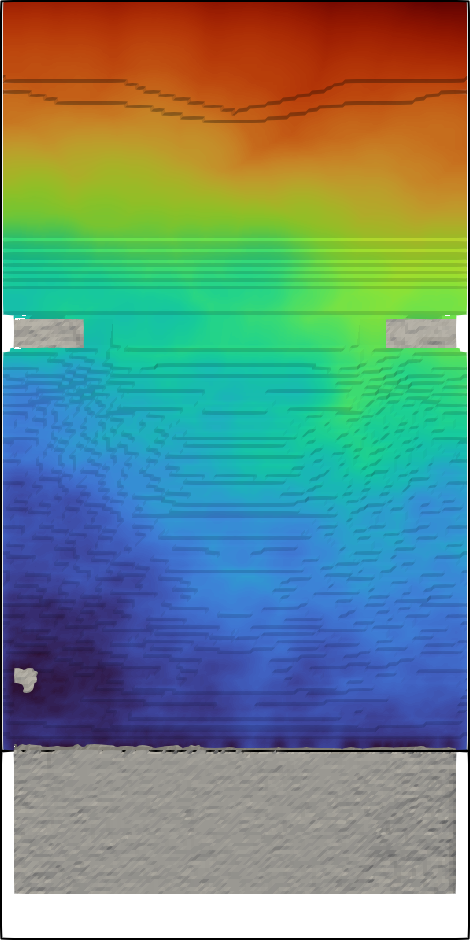}} \\
        {\footnotesize (left)} &{\footnotesize (front)} &{\footnotesize (right)}\\[-.5em]
    \end{tabular}}
    \subfloat[]{\label{fig:crack_surf_H45}
    \begin{tabular}{ccc}
\multicolumn{3}{c}{\footnotesize \textsf{H45} test, $\Delta {s}_{\text{max}} =$ 1.18~mm} \\
        \fbox{\includegraphics[height=3.5cm]{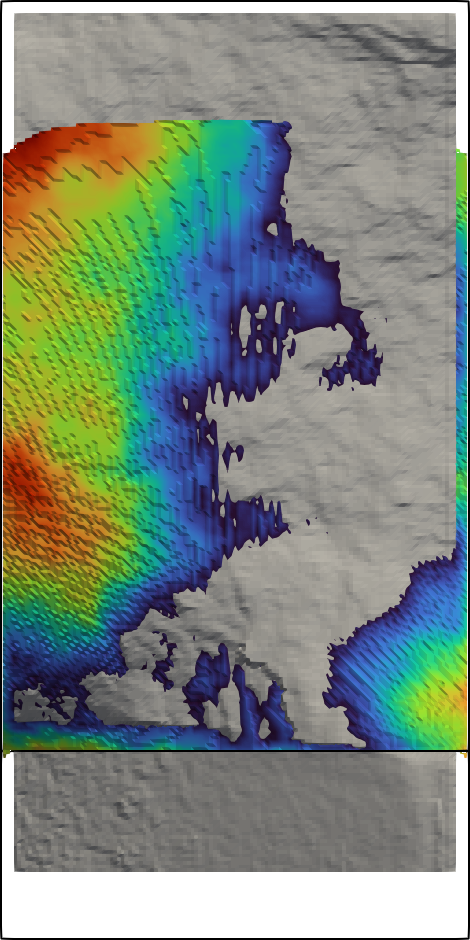}} & \fbox{\adjincludegraphics[height=3.5cm, trim={{.35\width} 0 {.35\width} 0}, clip]{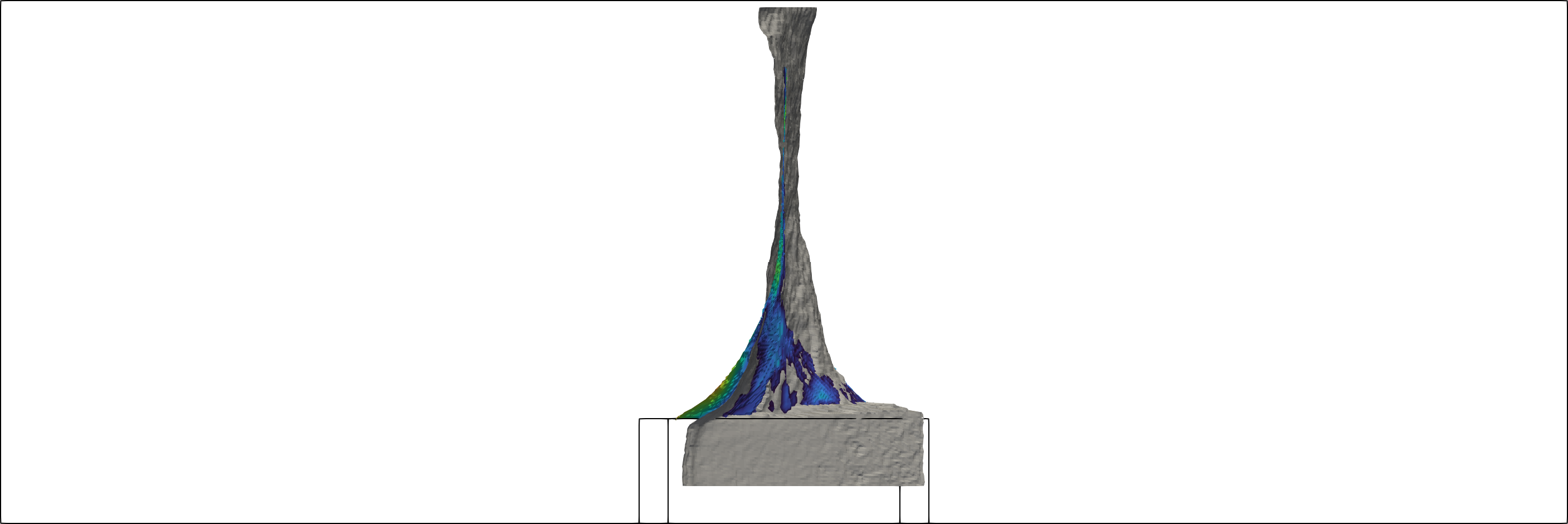}} & \fbox{\includegraphics[height=3.5cm]{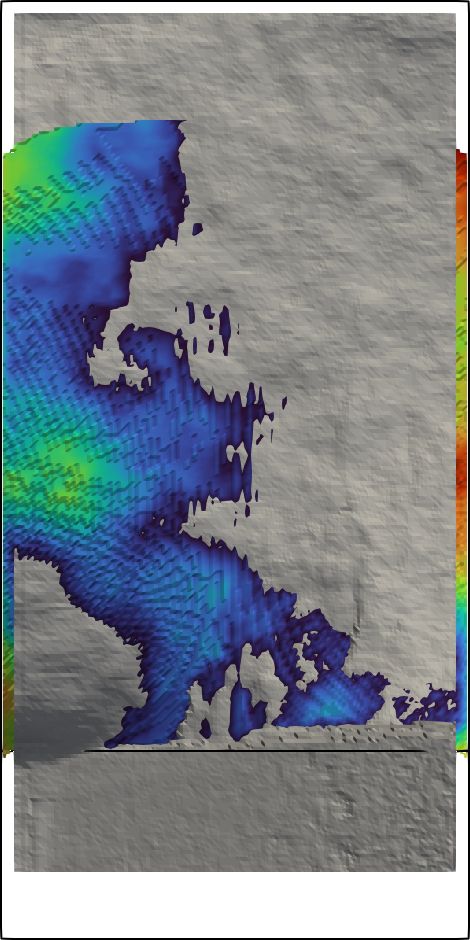}} \\
        {\footnotesize (left)} &{\footnotesize (front)} &{\footnotesize (right)}\\[-.5em]
    \end{tabular}}\\[0.1cm]
    \includegraphics{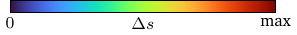}
    \caption{Comparison between numerical and experimental crack surfaces for the (a) \textsf{HC}, (b)  \textsf{HB}, (c) \textsf{HA} and (d) \textsf{H45} specimens. (Gray and colored surfaces represent experimental and numerical results, respectively).}
    \label{fig:calibration_crack_surfaces_images_surf}
\end{figure}

Fig.~\ref{fig:cracksurface_cuts_calibration} presents a comparison between the experimental and numerical crack paths within the three horizontal sections $A\sdash A$, $B\sdash B$ and $C\sdash C$ illustrated in Fig.~\ref{fig:crack_surfaces_images_thres}.
Here, the set of points describing for each test the numerical crack surface $\bs{s}_{\text{num}}$ and the experimental average surface $\bs{s}_{\text{exp}}$ along with the range over which they vary are obtained, as detailed in \ref{app:postprocessing}.
To obtain a quantitative evaluation of the agreement between numerical and experimental results, the shortest distance $\Delta s = \min\, \lVert{\bs{s}_{\text{num}}-\bs{s}_{\text{exp}}}\rVert$ is computed as deviation measure (see \ref{app:postprocessing}).
The agreement between experimental and numerical results is very good, with $\Delta s <$ 2~mm almost everywhere (Fig.~\ref{fig:cracksurface_cuts_calibration}).
The only exception is the $A\sdash A$ section of the \textsf{HA} test, where a maximum deviation of 4.68~mm is observed.
This is due to the homogeneous Dirichlet condition at the contact line between loading rod and specimen, which forces the crack to propagate away from it.
A complete comparison of the numerical and experimental crack surfaces is given in Fig.~\ref{fig:calibration_crack_surfaces_images_surf}, where we can appreciate that the model is able to predict with reasonable accuracy a wide range of fracture propagation modes under different degrees of mode mixity.
This is particularly evident from the \textsf{H45} specimen (Fig.~\ref{fig:crack_surf_H45}), for which the proposed model and calibration procedure deliver an accurate load-deflection curve (Fig.~\ref{fig:forcedisplacement_calibration}) as well as a good prediction of the complex crack path.

Augmented-reality (AR) renders of the final phase field (Fig. \ref{fig:crack_surfaces_images_thres}) along with a comparison of the processed experimental and numerical crack surfaces (Fig.~\ref{fig:calibration_crack_surfaces_images_surf}) can be accessed at \url{https://ar.compmech.ethz.ch} or using the QR-code in \ref{app:AR_render}.

\subsection{Blind prediction of the \textsf{DMC} test}\label{sct:blind_DMC}
\begin{figure}[!hbt]
    \centering\includegraphics{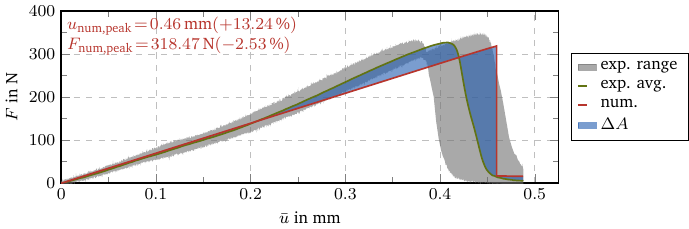}
    \caption{Comparison between numerical and experimental load-deflection curves for the \textsf{DMC} test.}\label{fig:forcedisplacement_challenge}
\end{figure}

After the calibration of the model (Section~\ref{sec:parameter_calibration}) and its verification (Section~\ref{sct:comp_calibr}), we finally present the results of the blind numerical prediction of the \textsf{DMC} test along with the experimental data available in \cite{dmc_challenge}.
The comparison between numerical and experimental load-deflection curves in Fig.~\ref{fig:forcedisplacement_challenge} demonstrates that the proposed approach is able to blindly predict the behavior of the DMC specimen with a very good accuracy, considering the experimental scatter.
The predicted peak force is very close to the average one obtained experimentally (with a deviation of $-$2.5~\%). For the displacement corresponding to the average peak force, the error is slightly higher but still reasonably low ($+$13.24~\%).
The displacement value at which the force drops almost to zero is very well captured (Fig.~\ref{fig:forcedisplacement_challenge}). 

\begin{figure}[!hbt]
    \subfloat[]{\label{fig:DMC_front}
         \centering
         \raisebox{0.09cm}{\includegraphics[width=7.125cm]{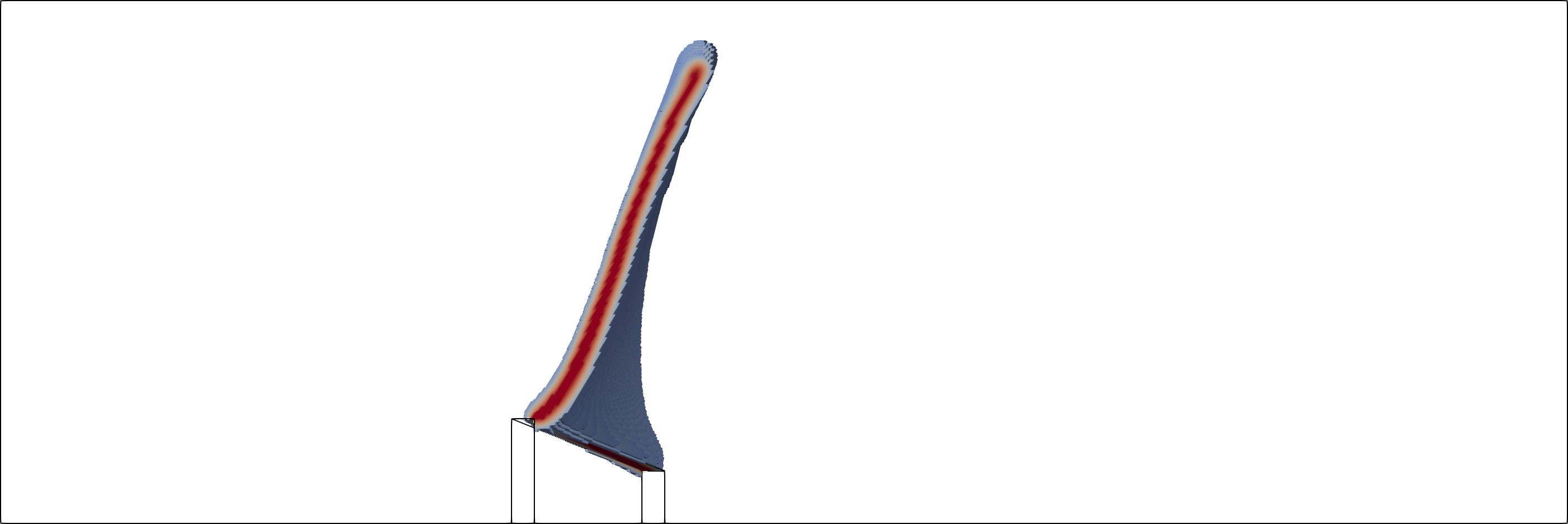}}
     }
    \subfloat[]{\label{fig:DMC_back}
         \centering
         \begin{tikzpicture}
             \footnotesize
             \draw[grau, thin, dash dot] (-0.5,0.9062) -- (1.5,0.9062); 
             \draw[grau, thin, -stealth] (-0.5,0.9062+0.25) -- (-0.5,0.9062);
             \draw[grau, thin, -stealth] (1.5,0.9062+0.25) -- (1.5,0.9062);
             \node[grau, anchor=east] at (-0.5,0.9062+0.125) {A};
             \node[grau, anchor=west] at (1.5,0.9062+0.125) {A};
             \draw[grau, thin, dash dot] (-0.5,0.0919) -- (1.5,0.0919); 
             \draw[grau, thin, -stealth] (-0.5,0.0919+0.25) -- (-0.5,0.0919);
             \draw[grau, thin, -stealth] (1.5,0.0919+0.25) -- (1.5,0.0919);
             \node[grau, anchor=east] at (-0.5,0.0919+0.125) {B};
             \node[grau, anchor=west] at (1.5,0.0919+0.125) {B};
             \draw[grau, thin, dash dot] (0.2,-1.12) -- (1.5,-0.57); 
             \draw[grau, thin, -stealth] (0.2,-1.12+0.25) -- (0.2,-1.12);
             \draw[grau, thin, -stealth] (1.5,-0.57+0.25) -- (1.5,-0.57);
             \node[grau, anchor=east] at (0.2,-1.12+0.125) {C};
             \node[grau, anchor=west] at (1.5,-0.57+0.125) {C};
             \node[anchor=center] at (0,0) {\includegraphics[width=7.125cm]{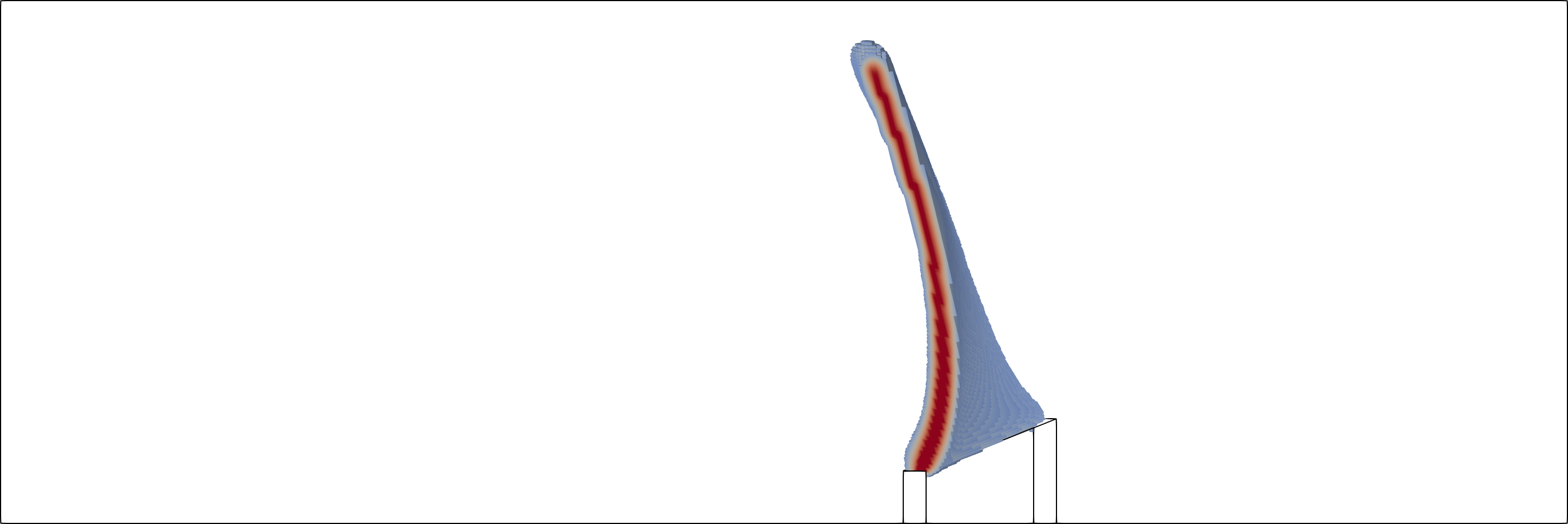}};
         \end{tikzpicture}}\\[-3.3cm]
       \subfloat[]{\label{fig:DMC_top}
         \centering
         \includegraphics[width=7.125cm]{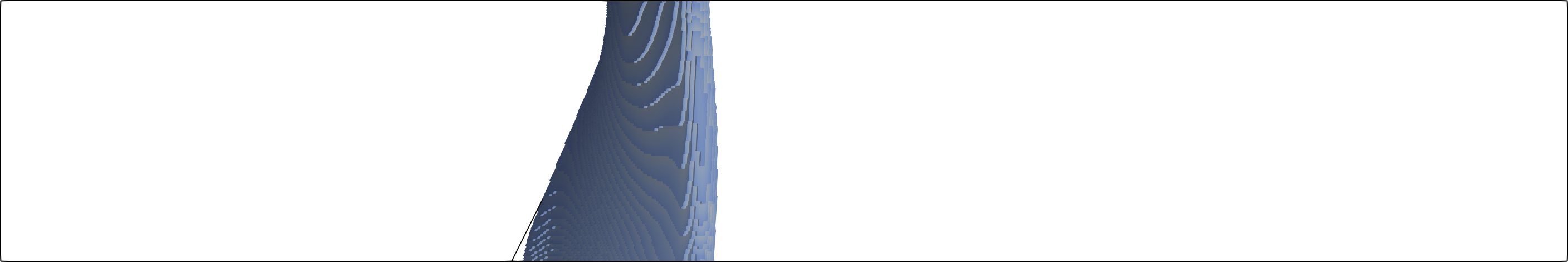}
     }
        \subfloat[]{\label{fig:DMC_bottom}
         \centering
         \includegraphics[width=7.125cm]{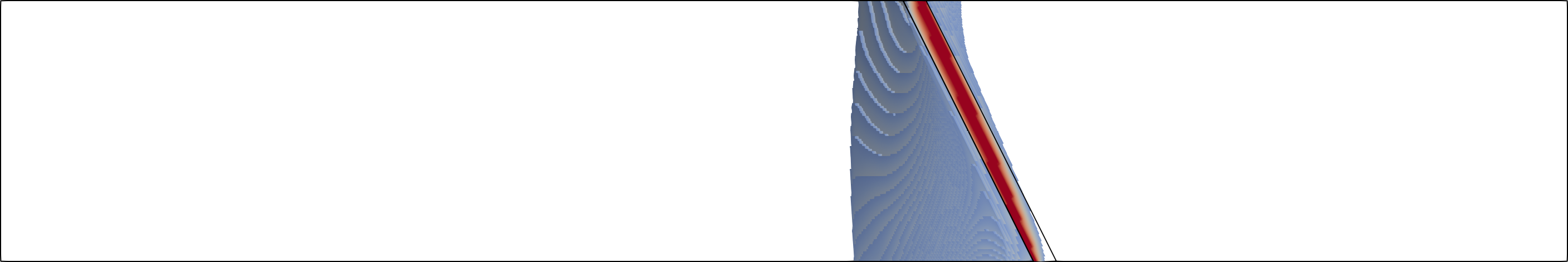}
    }
     \hspace*{0cm}
     \begin{subfigure}[b]{0.05\linewidth}
         \hspace*{0.5cm}\raisebox{0.6cm}{\includegraphics{colormap_pf.pdf}}
     \end{subfigure}
     \caption{Crack path at the last load step obtained numerically for the \textsf{DMC} specimen: (a) front, (b) back, (c) top and (d) bottom views.
     (All points with $d\le$ 0.25 are hidden for the sake of clarity).}
     \label{fig:challenge_crack_surfaces_images_thres}\setcounter{subfigure}{0}
 \end{figure}

We now present the numerically obtained crack path, displayed in Fig.~\ref{fig:challenge_crack_surfaces_images_thres} as the set of points with phase-field values $d\ge$ 0.25.
The model predicts a quite complex crack path, with a front that twists and bends in direction of the midsection as the crack propagates toward the upper surface. This complicated 3D path results from the mixed-mode fracture conditions with all three modes involved (I+II+III).
An AR 3D render of the phase-field crack surface can also be accessed at \url{https://ar.compmech.ethz.ch} or using the QR code in \ref{app:AR_render}.

\begin{figure}
    \centering
    \includegraphics{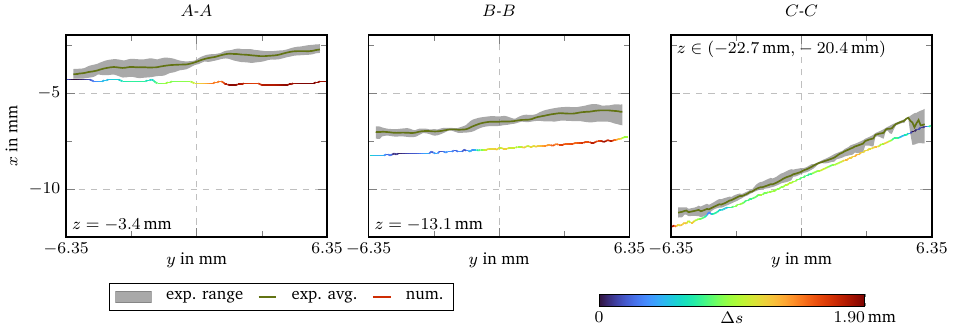}
    \caption{Comparison between numerical and experimental crack paths within the three horizontal sections $A\sdash A$, $B\sdash B$ and $C\sdash C$ illustrated in Fig.~\ref{fig:DMC_back}.}
    \label{fig:cracksurface_cuts_challenge}
\end{figure}

\begin{figure}[!hbt]
    \centering
    \begin{tabular}{cccc}
        \fbox{\includegraphics[height=3.75cm]{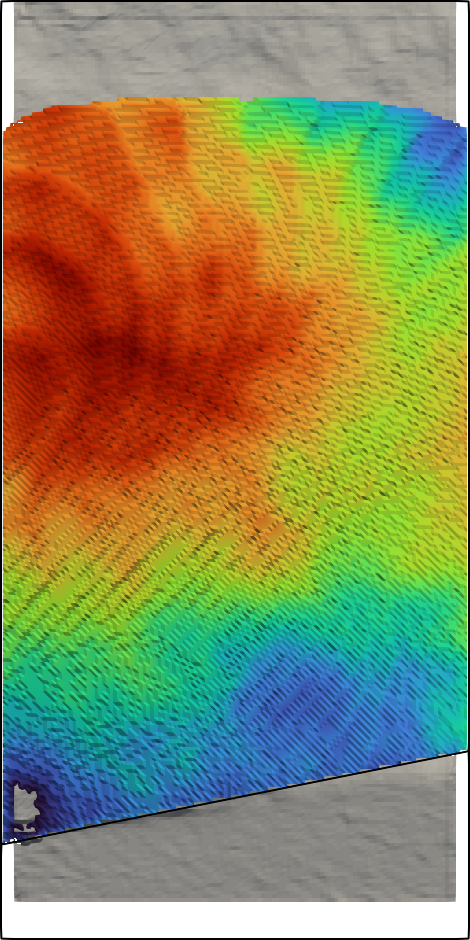}} & \fbox{\adjincludegraphics[height=3.75cm, trim={{.3\width} 0 {.3\width} 0}, clip]{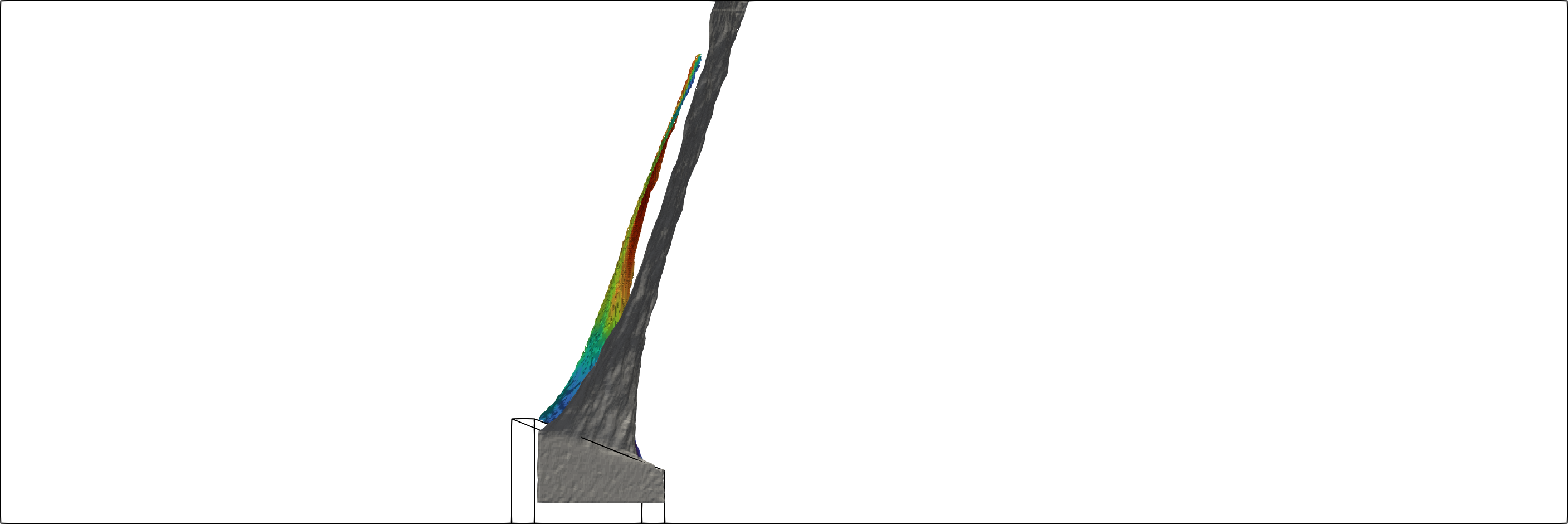}} & \fbox{\includegraphics[height=3.75cm]{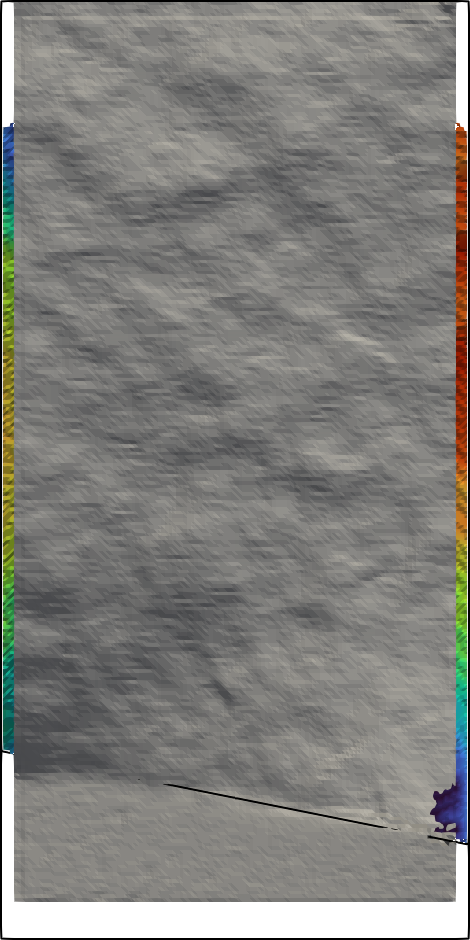}} & \fbox{\adjincludegraphics[height=3.75cm, trim={{.3\width} 0 {.3\width} 0}, clip]{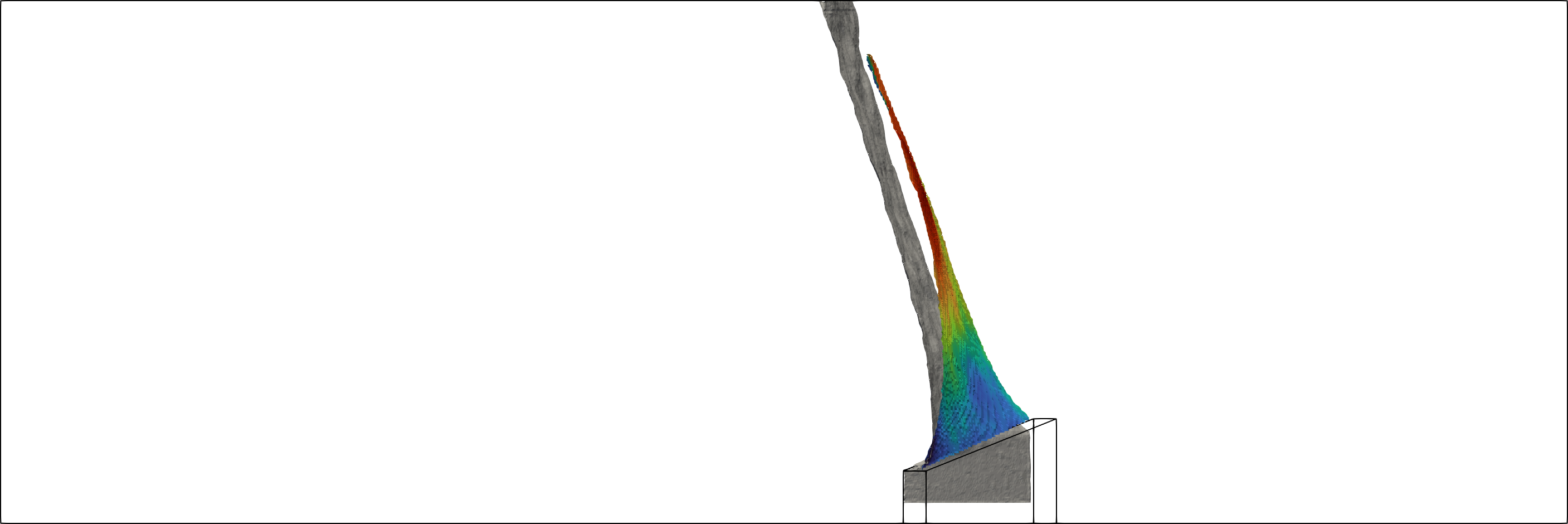}}\\
        {\footnotesize (left)} &{\footnotesize (front)} &{\footnotesize (right)} &{\footnotesize (back)}\\[-.8em]\\
    \end{tabular}
    \hfill\raisebox{-1.75cm}{\includegraphics{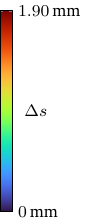}}
    \caption{Comparison between numerical and experimental crack surfaces for the \textsf{DMC} specimen.
    (Gray and colored surfaces represent experimental and numerical results, respectively).}
    \label{fig:challenge_crack_surfaces_images_surf}
\end{figure}

For the three horizontal sections $A\sdash A$, $B\sdash B$ and $C\sdash C$ (Fig.~\ref{fig:DMC_back}), Figs.~\ref{fig:cracksurface_cuts_challenge} and \ref{fig:challenge_crack_surfaces_images_surf} present a comparison between numerical and experimental crack paths, obtained similarly to what described in Section~\ref{sct:comp_calibr}.
Also in this case an excellent agreement is achieved, with a deviation $\Delta s$ always below 2~mm.
We remark that this result is particularly important since it is obtained with material parameters calibrated on a completely independent set of tests. 

\section{Summary and conclusions}\label{sec:conclusions}
The main objective of this work is the blind prediction of the behavior up to failure of a notched specimen, made of an artificial gypsum through additive manufacturing and tested in three-point bending. The notch geometry induces mixed-mode fracture conditions involving all three modes (I+II+III).
This specific task is defined in the context of the \textit{Damage Mechanics Challenge} issued by Purdue University  \cite{dmc_calibration1,dmc_calibration2,dmc_slides}.

To this end, a phase-field model for fracture in a brittle orthotropic material is calibrated using a set of provided experimental tests.
The calibration procedure involves two stages: first, we estimate most of the elastic properties of the orthotropic material using plane wave velocity and unconfined compressive test results; then, we perform an optimization process to identify the remaining elastic parameter and the fracture toughness of the material.
The optimization process is based on the minimization of a cost function encoding the mismatch between an independent set of experimental results on notched beams under three-point bending and the related numerical predictions.
To verify the calibrated model, we compare numerical and experimental results for the calibration tests in terms of load-deflection curves and crack paths. Finally, we compare the blind prediction of the challenge test results to the experimental results provided by the challenge issuer. Based on this work, the following conclusions can be drawn:
\begin{itemize}[-]
    \item the adopted calibration procedure is an effective way of deriving elastic and fracture parameters of the material from the diverse set of available test results, while keeping the computational cost of the inverse problem to a manageable level;
    \item considering the scatter of the experimental results and the many simplifying assumptions inherent to the macroscopic model at hand, the numerical predictions are in very good qualitative and quantitative agreement with the experimental results for both calibration and challenge tests, in terms both of load-displacement curves and of crack paths;
    \item the adopted phase-field model is able to quantitatively predict the propagation of cracks under different local stress conditions, including various degrees of mode mixity which may also result in complex and fully three-dimensional crack paths.   
\end{itemize}

\appendix
\setcounter{figure}{0}

\section{Crack surface post-processing}\label{app:postprocessing}
For the submission of the results in the DMC, the obtained numerical results need to be post-processed. In particular, the definition of the crack surfaces needs specific attention and a proper procedure.

\begin{figure}[t]
    \centering
    \subfloat[]{
    \begin{tikzpicture}
        \node at (-1.5,1.75) {\includegraphics{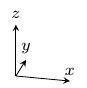}};
        \node at (0,0) {\includegraphics[scale=0.1]{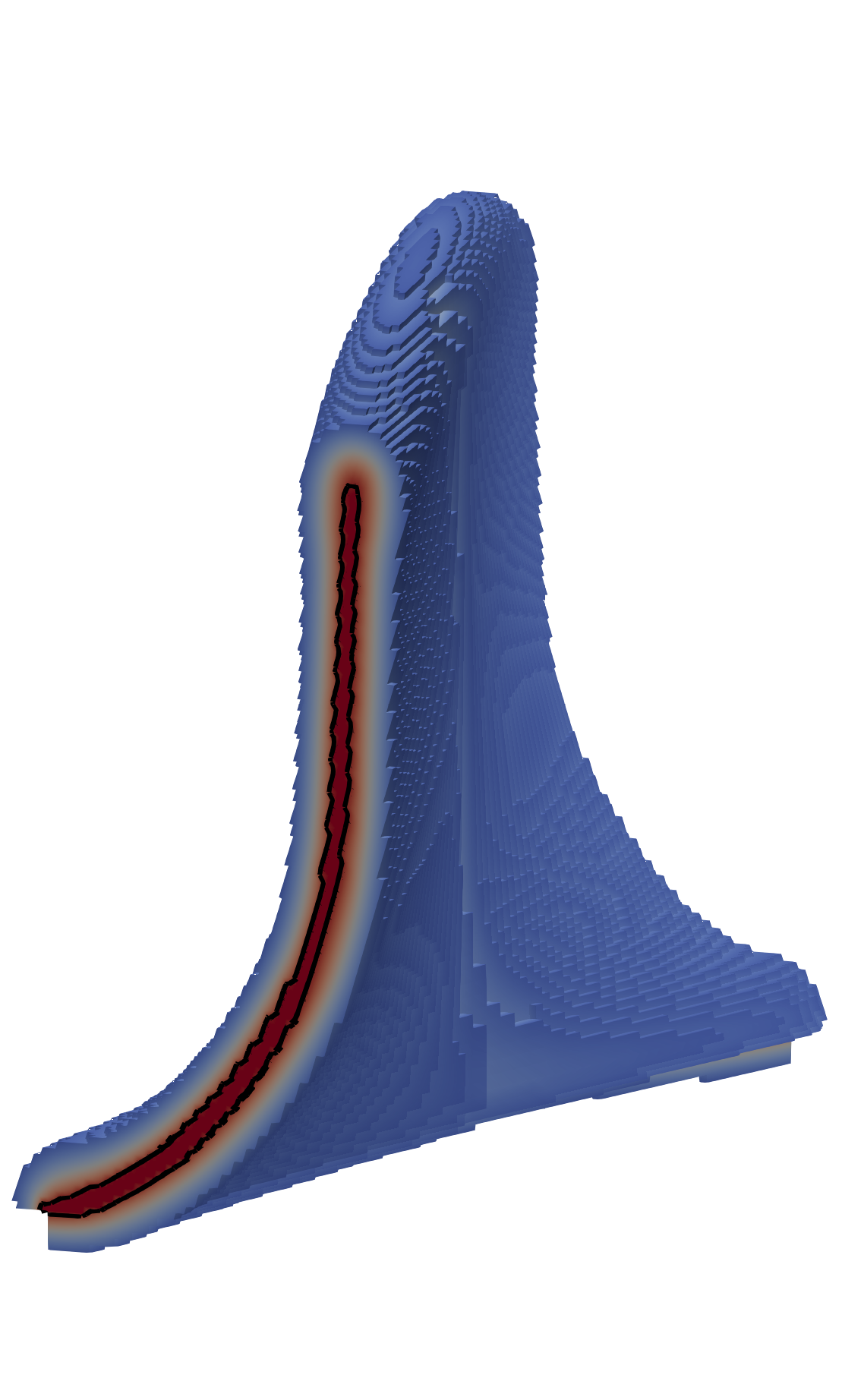}};
        \node[anchor=north east] at (2.75,2.4) {\includegraphics{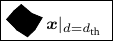}};
    \end{tikzpicture}
    \hspace*{-0.5cm}
    \label{fig:processing_crack_surface_phasefield}
    }
    \hspace*{0.25cm}
    \subfloat[]{
        \begin{tikzpicture}
            \node at (0,0) {\includegraphics[scale=0.1]{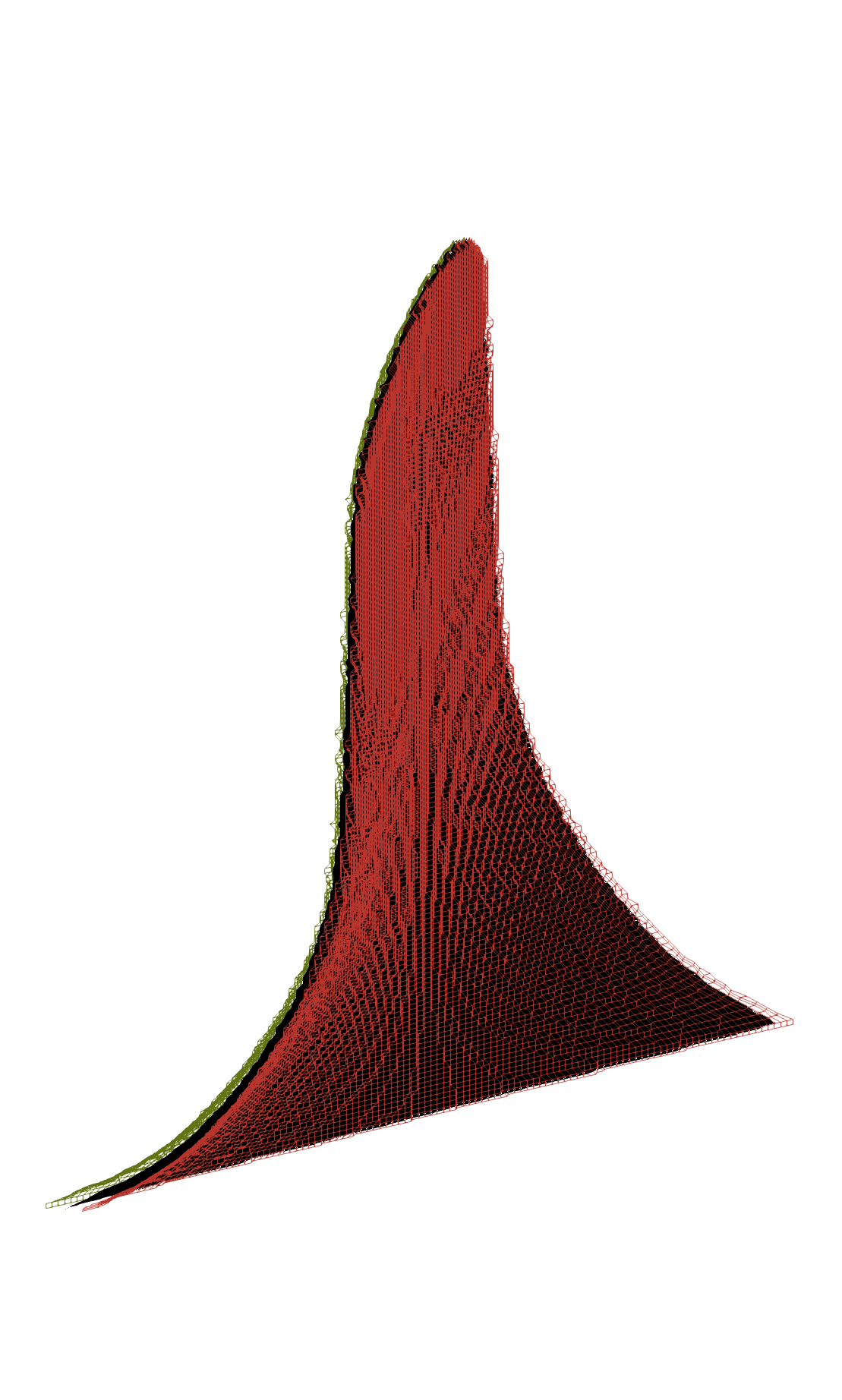}};
            \node[anchor=north east] at (2.75,2.4) {\includegraphics{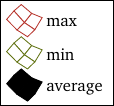}};
        \end{tikzpicture}
        \hspace*{-0.5cm}
        \label{fig:processing_crack_surface_numerical}
    }
    \hspace*{0.25cm}
    \subfloat[]{
        \begin{tikzpicture}
            \node at (0,0) {\includegraphics[scale=0.1]{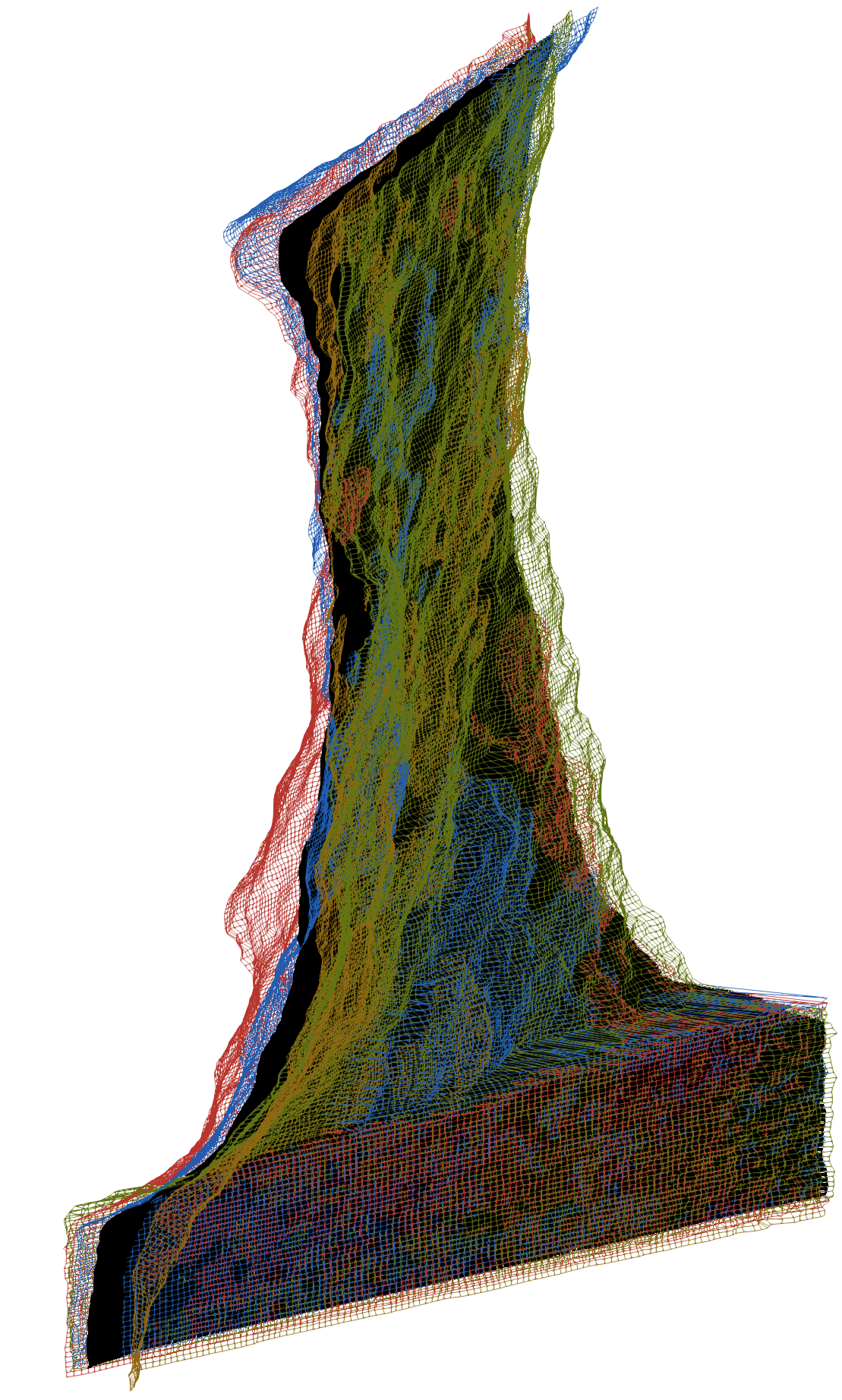}};
            \node[anchor=north east] at (2.75,2.82) {\includegraphics{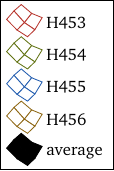}};
        \end{tikzpicture}
        \label{fig:processing_crack_surface_experimental}
    }
    \caption{(a) Crack volume (black) defined by the phase-field iso-surface with $d=0.95$ (all elements with $d<0.1$ are hidden for visibility).
    (b) Definition of the numerical crack surface from averaging the minimum (green) and maximum (red) crack surface coordinates with $d>0.95$.
    (c) Post-processing of the crack surface asperity measurements (colored) and average experimental crack surface (black), data from \cite{dmc_calibration2}.}
    \label{fig:processing_crack_surface}
\end{figure}

In this work, a material point is regarded as broken for a phase-field value above a threshold set to $d_{\text{th}} =$ 0.95.
This yields a crack volume enclosed into the 3D iso-surface with $d = d_{\text{th}}$ (Fig.~\ref{fig:processing_crack_surface_phasefield}).
This 3D iso-surface is obtained by applying the `Contour' filter to the numerical results in \texttt{ParaView} \cite{paraview1,paraview2}, which automatically creates a triangulation of the crack surface.
From this 3D surface, the numerical crack tip is extracted as the subset of the points enclosed in the crack volume having maximum height coordinate along the thickness direction (Fig.~\ref{fig:processing_crack_surface_phasefield}).

Regarding the crack surface, the requirement for the challenge submission is to present a sampling of the crack surface based on a uniform discretization with step size of 0.1~mm. To this end, the phase-field results $d(\bs{x})$ are first resampled with the required step size using the \texttt{ParaView} `Resample With Dataset' filter.
This resampled phase-field point cloud is then analyzed using an ad hoc script to extract the two outer limit crack surfaces within which the predicted crack surface lies (indicated by the red and green grid in Fig. \ref{fig:processing_crack_surface_numerical}).
These are used for the DMC submission, where the two sides of the crack surface are requested.
Concerning the comparison of the crack paths a unique surface is needed, for which here the average of the two limit surfaces is used, indicated by the black surface in Fig. \ref{fig:processing_crack_surface_numerical}.

Regarding the experimental data, to obtain a unique experimental crack surface out of the multiple tests performed, the experimentally obtained crack surfaces are averaged as well.
To this end, the crack surface asperity measurements of the broken TPB specimens provided in \cite{dmc_calibration2} are first cleaned from any offsets or outliers (either manually or automatically with the `Filloutliers' function of \texttt{Matlab} \cite{MATLAB}).
The corrected crack surface asperity measurements are then properly placed within the $x$- and $z$-direction of the geometry based on the coordinates of a significant point of an edge of the notch.
As illustrated exemplarily in Fig.~\ref{fig:processing_crack_surface_experimental} for the \textsf{H45} specimen, a representative experimental crack surface (black surface) is then identified with the average of the corrected and aligned crack surfaces of the four specimens (colored grids).

\section{Augmented reality renders}\label{app:AR_render}
\setcounter{figure}{0}

The AR 3D renders are created following the procedure presented in \cite{mathur_2023_brief}, and can be viewed on any recent mobile phone, tablet or internet browser.
Along with the comparisons of the crack surfaces, also AR renders of the final phase-field state (Fig.~\ref{fig:crack_surfaces_images_thres}, Fig.~\ref{fig:challenge_crack_surfaces_images_thres}) as well as the anisotropy visualizations (Fig.~\ref{eq:orthotropy}) are shown.

\begin{figure}[!hbt]
    \centering
        \includegraphics[width=4cm]{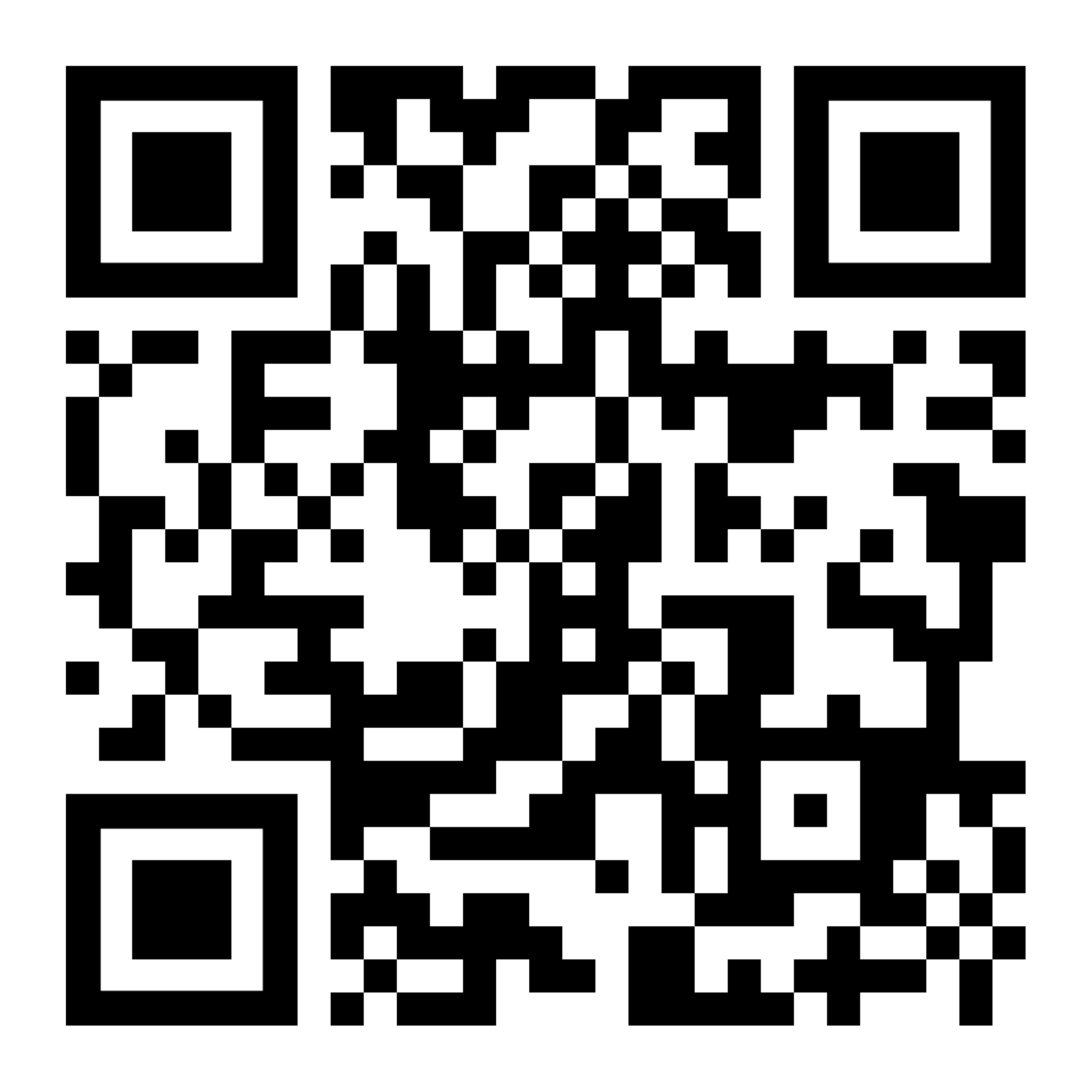}
    \caption{QR code for accessing the AR renders of the models.} \label{fig:ar_qr}
\end{figure}

The renders are accessible at \url{https://ar.compmech.ethz.ch} or using the QR-code in Fig.~\ref{fig:ar_qr}.
Further, all scripts used for an automatic model and website generation can be found at \url{https://gitlab.ethz.ch/compmech/ar}.

\section{Team background information}
Our team is comprised of doctoral students and postdoctoral researchers in the Computational Mechanics Group led by Prof. De Lorenzis at ETH Z\"urich.
The research of the group revolves around the development of solid mechanics models, along with novel techniques for their numerical solution and experimental validation.
At the moment, one of the group's main research topics is phase-field modeling of fracture and fatigue.
Previous collaborations of the team members on the phase-field approach to fracture drove the decision to explore the capabilities of the model applied to complicated 3D cracks in an anisotropic material within the Damage Mechanics Challenge.

\section*{Declaration of Interests}
The authors declare that they have no known competing financial interests or personal relationships that could have appeared to influence the work reported in this paper.

\section*{Acknowledgments}
We acknowledge funding from the European Union’s Horizon 2020 research and innovation programme under the Marie Skłodowska-Curie grant agreement No. 861061 – NEWFRAC Project, and from the Swiss National Science Foundation through Grant No. 200021-219407 `Phase-field modeling of fracture and fatigue: from rigorous theory to fast predictive simulations'.

\bibliography{references}
\bibliographystyle{elsarticle-num-names}

\end{document}